# Link between the Potentially Hazardous Asteroid (86039) 1999 NC43 and the Chelyabinsk meteoroid tenuous

Vishnu Reddy[1]
Planetary Science Institute, 1700 East Fort Lowell Road, Tucson, AZ 85719, USA
Email: reddy@psi.edu

David Vokrouhlický
Institute of Astronomy, Charles University, V Holešovičkách 2,
CZ-18000 Prague, Czech Republic

William F. Bottke
Southwest Research Institute, 1050 Walnut St, Suite 300, Boulder, CO 80302, USA

Petr Pravec
Astronomical Institute, Academy of Sciences of the Czech Republic, Fričova 1, CZ-25165 Ondřejov, Czech Republic

Juan A. Sanchez[1]
Planetary Science Institute, 1700 East Fort Lowell Road, Tucson, AZ 85719, USA

Bruce L. Gary
Hereford Arizona Observatory, Hereford, AZ, USA

Rachel Klima
Johns Hopkins University/ Applied Physics Laboratory, Laurel, MD, USA

Edward A. Cloutis
Department of Geography, University of Winnipeg, 515 Portage Avenue, Winnipeg, Manitoba, Canada R3B 2E9

Adrián Galád
Astronomical Institute, Academy of Sciences of the Czech Republic, Fričova 1, CZ-25165 Ondřejov, Czech Republic

Tan Thiam Guan
Perth Exoplanet Survey Telescope, 20 Lisle St, Mount Claremont, WA 6010, Australia.

Kamil Hornoch
Astronomical Institute, Academy of Sciences of the Czech Republic, Fričova 1, CZ-25165 Ondřejov, Czech Republic

Matthew R. M. Izawa

Department of Geography, University of Winnipeg, 515 Portage Avenue, Winnipeg, Manitoba, Canada R3B 2E9

Peter Kušnirák
Astronomical Institute, Academy of Sciences of the Czech Republic, Fričova 1, CZ-25165 Ondřejov, Czech Republic

Lucille Le Corre[1]
Planetary Science Institute, 1700 East Fort Lowell Road, Tucson, AZ 85719, USA

Paul Mann
Department of Geography, University of Winnipeg, 515 Portage Avenue, Winnipeg, Manitoba, Canada R3B 2E9

Nicholas Moskovitz[1]
Lowell Observatory, 1400 W Mars Hill Road, Flagstaff, AZ 86001

Brian Skiff
Lowell Observatory, 1400 W Mars Hill Road, Flagstaff, AZ 86001

Jan Vraštil
Institute of Astronomy, Charles University, V Holešovičkách 2,
CZ-18000 Prague, Czech Republic

[1]Visiting Astronomer at the Infrared Telescope Facility, which is operated by the University of Hawaii under Cooperative Agreement no. NNX-08AE38A with the National Aeronautics and Space Administration, Science Mission Directorate, Planetary Astronomy Program.

Pages: 51
Figures: 13
Tables: 3

**Proposed Running Head:** Chelyabinsk parent asteroid

**Editorial correspondence to:**
Vishnu Reddy
Planetary Science Institute
1700 East Fort Lowell Road, Suite 106
Tucson 85719
(808) 342-8932 (voice)
reddy@psi.edu

**Abstract**

We explored the statistical and compositional link between Chelyabinsk meteoroid and potentially hazardous asteroid (86039) 1999 NC43 to investigate their proposed relation proposed by Borovička et al. (2013). First, using a slightly more detailed computation we confirm that the orbit of the Chelyabinsk impactor is anomalously close to the asteroid 1999 NC43. We find ∼(1-3) × $10^{-4}$ likelihood of that to happen by chance. Taking the standpoint that the Chelyabinsk impactor indeed separated from 1999 NC43 by a cratering or rotational fission event, we run a forward probability calculation, which is an independent statistical test. However, we find this scenario is unlikely at the ∼($10^{-3}$ -$10^{-2}$) level. Secondly, we note that efforts to conclusively prove separation of the Chelyabinsk meteoroid from (86039) 1999 NC43 in the past needs to meet severe criteria: relative velocity ≃1-10 m/s or smaller, and ≃ 100 km distance (i.e. about the Hill sphere distance from the parent body). We conclude that, unless the separation event was an extremely recent event, these criteria present an insurmountable difficulty due to the combination of strong orbital chaoticity, orbit uncertainty and incompleteness of the dynamical model with respect to thermal accelerations. This situation leaves the link of the two bodies unresolved and calls for additional analyses. With that goal, we revisit the presumed compositional link between (86039) 1999 NC43 and the Chelyabinsk body. Borovička et al. (2013) noted that given its Q-type taxonomic classification, 1999 NC43 may pass this test. However, here we find that while the Q-type classification of 1999 NC43 is accurate, assuming that all Q-types are LL chondrites is not. Our experiment shows that not all ordinary chondrites fall under Q-taxonomic type and not all LL chondrites are Q-types. Spectral curve matching between laboratory spectra of Chelyabinsk and 1999 NC43 spectrum shows that the spectra do not match. Mineralogical analysis of Chelyabinsk (LL chondrite) and (8) Flora (the largest member of the presumed LL chondrite parent family) shows that their olivine and pyroxene chemistries are similar to LL chondrites. Similar analysis of 1999 NC43 shows that its olivine and pyroxene chemistries are more similar to L chondrites than LL chondrites (like Chelyabinsk). Analysis of the spectrum using Modified Gaussian Model (MGM) suggests 1999 NC43 is similar to LL or L chondrite although we suspect this ambiguity is due to lack of temperature and phase angle corrections in the model. While some asteroid pairs show differences in spectral slope, there is no evidence for L and LL chondrite type objects fissioning out from the same parent body. We also took photometric observations of 1999 NC43 over 54 nights during two apparitions (2000, 2014). The lightcurve of 1999 NC43 resembles simulated lightcurves of tumblers in Short-Axis Mode (SAM) with the mean wobbling angle 20º-30º. The very slow rotation of 1999 NC43 could be a result of slow-down by the Yarkovsky-O'Keefe-Radzievskii-Paddack (YORP) effect. While, a mechanism of the non-principal axis rotation excitation is unclear, we can rule out the formation of asteroid in disruption of its parent body as a plausible cause, as it is unlikely that the rotation of an asteroid fragment from catastrophic disruption would be nearly completely halted. Considering all these facts, we find the proposed link between the Chelyabinsk meteoroid and the asteroid 1999 NC43 to be unlikely.

1.   **Introduction**

Identifying the parent asteroid of meteorites in terrestrial collections (asteroid-meteorite linkage) is one of the ultimate goals of meteoritics. To reach that goal, dynamical models have been employed to link bolides to their source regions in the main belt or parent asteroids in the near-Earth asteroid (NEA) population based on their orbits. Bolide orbits are primary calculated based on video all sky camera networks and also surveillance footage.

On February 15, 2013, a 17-20 meter diameter asteroid entered the atmosphere over Chelyabinsk, Russia, and disintegrated in an airburst with an estimated energy of ~500±100 kilotons of TNT (Brown et al., 2013). Orbit for the meteor was calculated using video recordings, which suggested a pre-impact orbit consistent with its origin in the inner main belt (Borovička et al., 2013). Borovička et al. (2013) noted that the orbit of the Chelyabinsk bolide also showed similarities with a Q-type potentially hazardous asteroid (PHA) (86039) 1999 NC43 (Binzel et al., 2004). However, no direct spectral/compositional link has been made beyond the dynamical argument as is the case for 1999 NC43.

Several hundred fragments from the airburst were recovered including a >600 kg meteorite that was hoisted from the bottom of Lake Chebarkul. Laboratory classification of recovered fragments by Kohout et al. (2013) shows that Chelyabinsk is an LL5 ordinary chondrite with olivine and pyroxene as major mineral phases. The meteorite included a dark-colored fine grain impact melt/shock component, which is a significant portion (1/3) of the meteorite apart from light-colored lithology typical of ordinary chondrites. LL chondrites are the least abundant of the ordinary chondrites (which include H, L and LL chondrites) comprising ~10% of observed ordinary chondrite falls.

Laboratory spectral study of the recovered meteorites by Popova et al. (2013) suggested that the dark and light components had distinct spectral properties. They noted that absorption band depth comparison showed that some material in Chelyabinsk might be similar to H chondrites rather than LL chondrites. However, this result has not been confirmed by more in depth spectral and mineralogical studies by Kohout et al. (2013) and Reddy et al. (2014).

Kohout et al. (2013) measured the spectra, composition, density, porosity, and magnetic susceptibility of the Chelyabinsk meteorite. They concluded that compositionally the shock blackened/impact melt and LL5 chondrite lithologies are indistinguishable although their spectra and albedo varied. Bulk (3.32 g/cm$^3$) and grain densities (3.51 g/cm$^3$) of Chelyabinsk measured by Kohout et al. (2013) are also consistent with those of LL chondrites. The same study reported porosity values ranging from 1.5 to 11.4%. Unlike other LL chondrites, Chelyabinsk is reported to have more metallic iron, placing it between LL and L chondrites (Kohout et al., 2013).

Here we revisit in some detail statistical, dynamical and compositional arguments that led Borovička et al. (2013) to propose the link between 1999 NC43 and the Chelyabinsk meteorite: in Sec. 2 we present statistical and dynamical arguments, while in Sec. 3 we study the compositional similarity between 1999 NC43 and the Chelyabinsk body. Compositional links between Chelyabinsk and the source family in the main belt (Flora family) is the subject of another separate but related paper (Reddy et al., 2014).

## 2. Statistical and dynamical considerations

In this section, we examine whether the Chelyabinsk bolide could have came from 1999 NC43 using statistical and dynamical tools. Borovička et al. (2013) suggested a connection between the two bodies, with the Chelyabinsk bolide possibly ejected from 1999 NC43 via a collision. They based this on a brief probability calculation showing that the orbits of the bodies had a $\sim 10^{-4}$ chance of being similar to one another when compared to all known NEO orbits. Here we conduct a similar, though little more detailed calculation. We confirm that the orbit of the Chelyabinsk impactor is indeed anomalously close to the orbit of 1999 NC43, finding the probability is $\sim(1\text{-}3) \times 10^{-4}$ (Sec. 2.1). Adopting the hypothesis that the Chelyabinsk impactor is indeed a fragment ejected from 1999 NC43 in a significant cratering or mass-shedding event, we next perform a forward probability calculation that it would hit the Earth. While approximate, our result shows this is again an unlikely event with a probability of $\sim(10^{-3}\text{-}10^{-2})$ (Sec. 2.2). More support for the 1999 NC43-Chelyabinsk link hypothesis would come from a convergence of the two orbits in the past. However, the planet-crossing space is a dynamically harsh arena for this experiment to succeed: dynamical chaos caused by resonant interactions with the planets and the effects of planetary close encounters, unconstrained thermal accelerations and uncertainty in the Chelyabinsk pre-atmospheric orbit make it difficult to impossible to propagate NEO orbits far back into the past and thereby meet the conditions required by a collisional separation scenario. We give an example of such computation in Sec. 2.3.

### 2.1 Probability of the Chelyabinsk Orbit to be near that of 1999 NC43

Before we describe our next analyses, we revisit the method used by Borovička et al. (2013) to determine a statistical link between the bodies based uniquely on their current orbits. Recall that taking the derived orbit of Chelyabinsk, these authors looked for the NEO with the most similar orbit in the known population over all of the Keplerian orbital elements except mean anomaly: this was 1999 NC43 at a distance of $d \simeq 0.050$ using the Southworth and Hawkins (1963) metrics and $d \simeq 0.018$ using the Drummond (1981) metrics. They also found that 227 NEOs brighter than this object can be found among the known NEOs.

Since Borovička et al. (2013) do not give details of their method, we decided to first re-derive their probability analysis. We consider the Chelyabinsk impactor to be a single-class event with no equivalent over the past decades. In order to put its individual orbit into the context of a population of possible NEA impactors, we used the debiased model of Bottke et al. (2002). These authors determined likelihood to find a NEA in coarse semimajor axis $a$, eccentricity $e$, and inclination $I$ bins with volumes of 0.1 au × 0.05 × 5°, respectively. The Bottke et al. (2002) model was constrained to NEA sizes of approximately 375 m and larger, thus not extending all the way to Chelyabinsk size. Nevertheless, lacking a more advanced model, we use it as a good zero order approximation. Because the Chelyabinsk body impacted the Earth an additional weight over the Bottke et al. (2002) model should be based on the collision probability computation. Therefore, we recalibrated the likelihood function characterizing occupancy of the NEA orbits in the ($a,e,I$) space by multiplying it with a collisional probability from that particular orbital cell. We used an approximation based on Öpik-Wetherill theory from Bottke and Greenberg (1993). Using that scheme, we obtained a new probability distribution for a possible Chelyabinsk impactor orbit in the ($a,e,I$) space. Note that the collision probabilities increase especially in the regions where the orbital pericenter or apocenter becomes close to mean heliocentric distance of the Earth orbit, or for orbits very similar to the Earth (i.e., semimajor axis close to 1 au, and small value of eccentricity and inclination; see Fig. 5 in Morbidelli and Gladman, 1998).

To proceed toward the orbital distance computation, we must also assign two secular angles, longitude of node $\Omega$ and argument of pericenter $\omega$, to the modeled orbit. We chose $\Omega$ at random in the available interval of values between 0° and 360°. As far as the value of $\omega$ is concerned, we randomly chose one of the four possible impact configurations determined by the ($a,e,I$) values. This procedure is again approximate, as it neglects effects of Kozai regime at higher inclinations and eccentricities for the collisional probability computation (e.g., Vokrouhlický et al., 2012), and also related non-uniformities in the distribution of orbital angles distribution (e.g., JeongAhn and Malhotra, 2013).

We used the above outlined procedure to create a large number of Earth impacting orbits from the NEA population. Each time we had an impactor orbit, we also created a set of 227 random large asteroids in the available NEA orbital space; these should model the 1999 NC43 analogs. Because 1999 NC43 did not impact the Earth, we obviously did not use the collisional probabilities in this case but rather directly the debiased likelihood distribution from Bottke et al. (2002) model. While not exact —note the Bottke et al. (2002) model orbit distribution was mostly constrained by asteroids with sizes between 0.5-1 km— this is a fairly good approximation. Use the debiased orbital distribution is also reasonable, because our knowledge of the population of D>2 km size asteroids population is basically complete. As far as secular angles $\Omega$ and $\omega$ are concerned for the large objects, we assumed uniform distribution in the 0° and 360° interval of values, again a fairly good approximation.

In each of many trials of drawing an Earth impacting body (aka Chelyabinsk body) and 227 large asteroids (analogs of 1999 NC43) from the simulated population of bodies in the dynamical space, we computed their orbital distances and determined their minimum value. We used both Southworth-Hawkins and Drummond orbital distances functions, and millions of trials. We then determined a fraction of cases in which the minimum distance was smaller than the critical value of the observed distance between the Chelyabinsk orbit and that of 1999 NC43. We obtained $\sim 3 \times 10^{-4}$ using the Southworth-Hawkins metrics and $\sim 0.8 \times 10^{-4}$ using the Drummond metrics. These results are in a very good agreement with those stated in Borovička et al. (2013).

We also repeated the previous procedure by taking the 1999 NC43 analogs from only the Earth-crossing space, rather than from the whole NEA population.

Note the latter also contains the Amor population of bodies not presently crossing the Earth's orbit. A priori the results may not be the same because the impactors and large bodies are now drawn from the same orbital-space zone making their orbits possibly closer to each other. On the other hand, there are fewer large (i.e., $D>2$ km) bodies in the Earth-crossing space. In particular, we find there are about 110 such bodies. Running the statistical analysis again we found the two factors compensated and the probability results were nearly the same as mentioned above. Therefore, independent of many of the details of the method, we recover the finding of Borovička et al. (2013), namely that the probability for the Chelyabinsk impactor orbit to be at the observed distance from a large asteroid 1999 NC43 by chance is low, $\sim(1\text{-}3) \times 10^{-4}$.

*2.2    Forward Likelihood Computation for the Chelyabinsk and 1999 NC43 Link*

Results described in Sec. 2.1 justify the hypothesis that the Chelyabinsk body and asteroid 1999 NC43 are dynamically related, possibly the former being a fragment produced during a large cratering or mass-shedding event in the past. To further test this possibility, we conducted a forward probability computation to evaluate the likelihood of this scenario. Note this is an independent statistical test from that presented in Sec. 2.1.

The probability we intend to derive is composed of two independent components: (i) probability that one of the large NEAs, D>2 km size category, experienced large cratering event over a certain time interval in the past, and (ii) probability that one of the Chelyabinsk-size fragments created in (i) hit the Earth within a past century or so. We call these individual probabilities $p_i$ and $p_{ii}$; since the processes in (i) and (ii) are uncorrelated, the combined probability is their product $p_i \times p_{ii}$.

In order to assess $p_i$, we must first estimate the timescale of interest over which we seek the possibility of the cratering event. To that goal we considered the current orbit of 1999 NC43 and created 100 orbital clones according to the covariance matrix from its orbital determination in the 6-dimensional space of orbital elements. Note that the dispersion of these clones about the nominal orbit of 1999 NC43 correspond to ~1 m/s ejection velocity in accordance with our estimate of reasonable ejection speeds. We then propagated all bodies forward in time using the swift code (e.g., http://www.boulder.swri.edu/~hal/swift.html). Every 10 years we checked the orbital configuration of the clones and evaluated their distance from the position of the nominal orbit of 1999 NC43 at that moment. We observed that the median orbital distance of the clones, both in Southworth-Hawkins and Drummond metrics, reaches the critical value of Chleyabinsk-1999 NC43 in 20,000 to 60,000 years. At 80,000 years, only 10% of the clones have distance smaller than Chelyabinsk to 1999 NC43. We verified that the results also hold if we include the thermal accelerations for the clones with values typical for 20 m bodies (e.g., Bottke et al., 2006). For simplicity we take 50,000 years as a characteristic timescale to disperse a stream of fragments beyond the nominal distance between the Chelyabinsk-1999 NC43 couple, and call it the decoherence timescale. Note our finding is in good accord with other similar studies (e.g., Pauls and Gladman, 2005).

Therefore, while this decoherence timescale was obtained for the stream dispersal around the orbit of 1999 NC43 only, it is reasonable to apply it to the full population.

Next, we must estimate the characteristic timescale for the 1999 NC43 clones to undergo a significant cratering event in the Earth-crossing zone. No specific study with that goal has been conducted to the best of our knowledge, so we use results from Bottke et al. (2005) who studied collisional processes in the main belt. In particular, they found that the collisional lifetime of a ~2 km size body is ~ 100 million years. We consider this value holds for large NEAs as well, though it is likely an underestimate, because only about half of the bodies cross the whole zone of the main belt, to seek potential projectiles, and even if doing so large inclination values would bring them high above the ecliptic plane. Because a large cratering event as opposed to collisional disruption is what we seek, we estimate that the characteristic timescale is about one tenth of the collisional lifetime. Therefore, we use ~10 million years as the characteristic timescale to produce a significant cratering event on large NEAs. Considering all the above, we readily obtain an estimate of $p_i \sim 0.55$. Therefore, of the estimated ~ 110 large NEAs in the Earth crossing zone, we conclude that approximately one of them might have undergone large cratering within the past ~ 50,000 years.

In order to estimate $p_{ii}$ (the probability that a fragment will impact Earth), we need the following information: (i) characteristic number of D ~ 20 m fragments created in a cratering event, and (ii) their collisional probability with the Earth. To estimate the first, we consider results in Durda et al. (2007). These authors evaluated a variety of impact geometries and projectile/target size ratios using a collisional simulation using smoothed-particle hydrodynamics (SPH) code. Importantly for us, their simulations also covered cratering events. Considering higher impact velocity cases, which should be appropriate for the impacts on NEAs (note that typical impact velocity of 1999 NC43 versus main belt projectiles is ~ 9 km/s), we estimate there might be about 10,000 Chelyabinsk-size fragments created. These alone represent an equivalent of ~ 430 m body which gives a rough impression of the event.

Next, we considered the orbits of the ~110 large asteroids in the Earth-crossing zone and determined their intrinsic collision probabilities with the Earth. To keep things simple we again used Opik-Wetherill scheme that assumes long-term constant values of the orbital eccentricity and inclination and performs collisional probability averaging over the secular cycles in angular variables (characteristic timescale of ~30,000 years). We found a median value of ~ 30 × 10$^{-18}$ km$^{-2}$ y$^{-1}$, with only 10% of values larger than ~ 100 × 10$^{-18}$ km$^{-2}$ y$^{-1}$. For reference, the intrinsic collision probability of the 1999 NC43 orbit is ~ 80 × 10$^{-18}$ km$^{-2}$ y$^{-1}$, at the larger end of the distribution. Based on this model the probability $p_{ii}$ will be ~ (30 × 10$^{-18}$ × 6400$^{\wedge}2$ × 10$^4$ × 100) ~ 1.3 × 10$^{-3}$; here 6400 km stands for the radius of the Earth and 100 y is the time window we expect the collision to happen. If then combined with $p_i \sim 0.55$, we would have the total probability of the scenario ~ 10$^{-3}$.

A possible caveat of the previous analysis may stem from using the collision probability computed as an average over secular cycle of the longitude of node Ω and argument of pericenter ω. While the former does not present a problem for the near-circular orbit of the Earth, the latter warrants some discussion. This is because the time-window over which we wish to characterize the impact probability, 100 years, is shorter than the typical timescale over which ω circulates. A characteristic population value for the latter may be the taken from the orbit of 1999 NC43, whose argument of pericenter circulates in ∼ 20,000 years. Recall, the collision probability is given by time spend near several impact-resulting values of ω divided by the reference timescale (e.g., Öpik, 1951; Wetherill, 1967). For orbits which potentially cross the impact configuration over the last century, this results in an increase of the collision probability by a factor ∼ 20000/ (4 × 100) = 50. The factor 4 comes from the 4 typical impact configurations during the secular cycle of ω, which is the minimum number (e.g., Vokrouhlický et al., 2012). Note that this estimate correctly takes into account that the impact-configuration window in ω (∼ 0.03° for 1999 NC43) is smaller that its change in 100 years (∼ 1.8° for 1999 NC43). Given that larger intrinsic collision probability of 1999 NC43 compared to the population median, we may assume that the collision probability increases by a factor of ∼100. Note that the orbit of 1999 NC43 is only about 14° from the collision orbit in ω, (the difference in Chelyabinsk orbit and that for 1999 NC43 (e.g., Borovička et al., 2013). So it is conceivable that fragments released from this asteroid within the past ∼ ($10^3$ – $10^4$) years might have advanced the necessary 14° to reach the collision orbit today, and thus the increased collision probability would apply to this orbit. This analysis would apparently increase the combined probability $p_i$ × $p_{ii}$ to about 0.1. However, this is not exactly true. We find that out of ∼110 large NEAs only about 10% have ω values close enough to the impact configuration with the Earth so that a more realistic estimate is $p_i$ × $p_{ii}$ ∼ 0.01.

Finally, we note that so far in this Section we assumed a cratering scenario as the mechanism for Chelyabinsk origin. If the ejection was due to a rotationally-driven mass-shedding, the odds would be even smaller. This is because the characteristic timescale to reach the fission limit due to the thermal torques for a *D* ∼ 2 km body is about 1-5 million years (e.g., Čapek and Vokrouhlický, 2004), only about an order of magnitude smaller than the characteristic timescale of the cratering event. However, the amount of Chelyabinsk-size bodies produced during such event would have been much smaller than in the cratering case. The presently slow rotation rate of 1999 NC43 would also require separation of a larger fragment (e.g., Pravec et al., 2010), that presumably might have been accompanied with a limited number of smaller ones.

*2.3 Difficulties of the Past Orbital Convergence in the NEO Zone: An Example of 2012 VV93 and 2013 CY10*

The statistical arguments in Secs. 2.1 and 2.2 do not provide a consistent picture. This does not mean they are incorrect, only that some assumptions may not be exactly satisfied on either side, and/or that fluke events happen. To further resolve the situation requires additional evidence. A strong argument can come from the past orbital evolution of the Chelyabinsk and 1999 NC43 orbits with the goal to seek their convergence and/or minimum conditions under which the Chelyabinsk body might have separated from that of 1999 NC43. Obviously, Borovička et al. (2013) were aware of the importance of this problem and they provided important initial dynamical constraints. In particular, they observed that the orbital similarity between the bodies persisted over 2,000 years, and that the ejection velocity of the Chelyabinsk body from 1999 NC43 was likely between $\simeq 0.7$ km/s and $\simeq 2$ km/s. While we do not question the accuracy of the calculation, we believe the interpretation is flawed, and that the ejection velocity predicted is anomalously high for typical collisional or rotationally-induced mass-shedding ejecta, and thus their experiment is inconclusive.

Lessons gleaned from the analysis of the small asteroid families in the main belt, which were created by catastrophic collisions or cratering events, indicate that typical dispersal velocities of the fragments are of the order of escape velocity from the parent object (e.g., Nesvorný et al., 2006; Nesvorný and Vokrouhlický, 2006). In the putative case of a body in the size range of 1999 NC43, this would imply separation velocities between 1-2 m/s. Even smaller relative velocities are observed for fragment separation by a rotational fission process that provides the population of observed asteroid pairs in the main belt (e.g., Vokrouhlický and Nesvorný, 2008, 2009; Pravec et al., 2010; see also Jewitt et al. 2014). In this case, the typical separation velocities of fragments from km-sized parents are between 0.01 m/s and 0.5 m/s.

Additionally, to build a convincing case for past separation for two bodies from a common parent body, it is necessary to bring them within a Hill sphere distance of one another by integrating their orbits backward in time. Again, in the case of the Chelyabinsk-1999 NC43 couple, this would mean a separation of $\simeq 100$ km. These convergence criteria, while highly restrictive, have been met by orbital configurations of asteroid pairs in the main belt (e.g., Vokrouhlický and Nesvorný, 2008, 2009; Pravec et al., 2010). This calculation is possible because orbital stability in the main belt zone is much higher than in the planet-crossing region. For the Chelyabinsk-1999 NC43 couple, the odds of convincingly proving past convergence of two NEOs orbits toward the separation point from a common parent body are very low unless the event is extremely young.

In order to gain insight into the problems of past convergence in the NEO zone, we look to the past history of NEOs 2012 VV93 and 2013 CY10. Not only are the current orbits closer than the Chelyabinsk-1999 NC43 couple (their orbital distance is $d \simeq 0.01$ using the Southworth and Hawkins metrics and $d \simeq 0.005$ using the Drummond metrics), but the expected orbital chaos for their orbits is smaller. This is because their current minimum orbital intersection distance (MOID) with the Earth is $\geq 0.2$ au. By definition, the Earth MOID of the Chelyabinsk orbit and 1999 NC43 is very small ($\leq 0.02$ au for 1999 NC43). This means that these orbits will potentially suffer large perturbations by Earth flybys, triggering an extreme divergence of close-by orbits. If the past orbital convergence in the NEO population has any possibility of success, it is in the orbital regions that are distant from the strong perturbations of Earth and Venus.

In the case of 2012 VV93 and 2013 CY10, neither have approached these planets in the near past. They belong to the Amor population, a sub-population of NEOs that are on solely Mars-crossing orbits. The problems here are (i) the poorly determined orbit of 2013 CY10 and (ii) its small size (not much larger than the estimated pre-atmospheric size of the Chelyabinsk bolide), which means the unconstrained thermal accelerations affecting its orbit are very large. Note that both are actually the same as in the case of Chelyabinsk pre-atmospheric orbit.

Here we use the methodology described in Vokrouhlický and Nesvorný (2008, 2009) and in the Supplementary information of Pravec et al. (2010). The two asteroids were represented by a large number of close clones expressing their current orbital uncertainty (we use 80 of them for 2012 VV93 and 160 of them for 2013 CY10). Each of these clones represented a certain number of asteroid "realizations" that were subject to different strengths of the Yarkovsky thermal accelerations (e.g., Bottke et al., 2006). Again we used 80 for 2012 VV93 and 160 for 2013 CY10 to account for these different accelerations. Altogether we had 6,400 and 25,600 possible past realizations of 2012 VV93 and 2013 CY10, respectively. We propagated this multitude of clones for 3,000 years into the past. At every 0.25 years we randomly selected 100 million possible identifications of the pairs between the clones of the first and second bodies and examined their distance and relative velocity. We sought to find configurations that would indicate a possible separation event.

The best results were for a separation of $d \simeq 20,000$ km and a relative velocity $V \simeq 20$ m/s. These values are one to two orders of magnitude higher than needed to claim a successful convergence result. Either the two bodies were separated in a more distant past, or more clones are needed to overcome the strong orbital divergence of the bodies even on such a short timescale as several kys. The third possibility, which we consider the most likely, is that the two bodies are unrelated. Regardless, these test cases show the difficulties in proving a dynamical link between two bodies in NEO space.

## 3. Physical Characterization

Discussion in Sec. 2.3 suggests there are difficulties in proving the orbital convergence between the Chelyabinsk pre-atmospheric orbit and that of 1999 NC43 in the past, using dynamical simulations. What remains then are physical studies of the two bodies. This is the topic of the current section in which we focus on the asteroid 1999 NC43. Before we present our analysis, we briefly review other available information about this asteroid (several facts about the Chelyabinsk body were outlined in Sec. 1).

1999 NC43 is an Apollo type near-Earth asteroid. Delbó et al. (2003) estimated a diameter of 2.22 km based on Keck thermal infrared observations. They derived an albedo of 0.14 based on Near-Earth Asteroid Thermal Model (NEATM), which is on the lower end of the range for Q-type asteroids. Due to uncertainties in the H magnitude, they estimated that the albedo could be 0.23 based on NEATM if one assumes H of 15.5 (Delbó et al., 2003). Based on the beaming parameter as a function of observed solar phase angle, Delbó et al. (2003) noted that 1999 NC43 has *"exceptionally high surface thermal inertia comparable to, or exceeding, that of solid rock."* More recent observations by the Akari spacecraft (Usui et al. 2011) suggest a diameter of 1.43 km, which would give an albedo of ~0.29 (consistent with Q-type asteroids). De León et al. (2010) analyzed near-IR spectrum of 1999 NC43 obtained by SMASS/MIT-UH-IRTF Joint Campaign for NEO Reconnaissance using methods described in Cloutis et al. (1986). They concluded that the taxonomy classification of 1999 NC43 is Q-type consistent with Binzel et al. (2004).

*3.1  Rotational Properties*

*3.1.1 Photometric Observations*
We took photometric observations of 1999 NC43 with the 0.65-m telescope in Ondřejov (11 nights in 2000 and 6 nights in 2014), the 1.54-m Danish telescope on La Silla (11 nights), the 0.3-m Hereford Arizona Observatory (16 nights), the 0.7-m telescope at Lowell Observatory in Arizona (7 nights), and 0.3-m T.G. Tan observatory in Perth, Australia (3 nights). The individual observing sessions and their observational circumstances are listed in Table 1. The mid-time (UTC) of the run, rounded to the nearest tenth of day, is given in the first column. Filter name, observation location and telescope size are given in subsequent columns.

The observations with the 0.65-m and the 1.54-m Danish telescopes were taken with the Bessell R filter, with supplementary observations in the V filter on 2014 April 4, and they were calibrated in the Johnson-Cousins system using Landolt (1992) standard stars. Integration times were 90-120 and 30-40 seconds, respectively, and the telescopes were tracked at half-apparent rate of the asteroid. Because of the asteroid's long period we did not need to take continuous observations, but we took a short series of typically five images twice a night, with a typical separation of a couple hours, depending also on scheduling constraints of our other asteroid observations we ran on the nights; we worked 1999 NC43 as a secondary target on most of the nights. The observations were reduced using procedures described in Pravec et al. (2006).

The observations with the 1.54-m and the 0.65-m telescopes were calibrated in the Johnson-Cousins VR system with absolute errors 0.01 to 0.02 mag. The observations from the other three stations were calibrated in the Sloan r' system with absolute errors of 0.03 mag. We combined them with the Cousins R data using r' = R + 0.19, with the offset derived from the overlapping data taken with the 1.54-m and the 0.7-m on 2014-03-26.2.

To homogenize the data taken with different sampling rates we averaged measurements taken from a single station on nearby times over a time span not longer than 0.5 h (0.4 % of the asteroid's period); typically five consecutive data points were averaged. We suppressed averaging more than nine measurements. The final dataset of Cousins R data consisting of 23 observations (normal points) from 2000, and 108 normal points from 2014, is available from the authors upon request.

*3.1.2 Photometric Analysis*
First, we run a single-period search using the Fourier series fitting technique (e.g., Pravec et al., 1996), restricted to the $2^{nd}$ order. For the analysis we reduced the data to unit geo- and heliocentric distances and to a consistent solar phase using the H-G phase relation, assuming the slope parameter G = 0.24 ± 0.11 that is the range for SQ types (Warner et al., 2009). While the 2000 data gave an ambiguous solution with two possible periods of 34.5 and 122 h, the 2014 data provided a unique period of 122.2 h, assuming a lightcurve dominated by the 2nd harmonic (i.e., with two pairs of maxima and minima per period). An uncertainty of the period determination is a few times 0.1 h, with the synodic-sidereal correction amounting up to 2 hours. In Fig. 1 we plot the reduced $\chi^2$ versus period; there, we conservatively assumed the error of 0.03 mag for all the points. It shows that even for the best fit period of 122.2 h (and its half of 61.1 h, which would be for a monomodal lightcurve with one maximum and minimum per period), the fit is poor and the mean residual of the fitted Fourier series is about 5 times larger than expected.

In Figures 2 and 3 the photometric data are folded with the period of 122.2 h. A detailed check of the 2014 data reveals a characteristic scatter pattern of the measurements. While a part of the scatter may be due to changes of viewing and illumination aspect during the 2-month long observational interval, a comparison of several groups of data points covering same phases on consecutive cycles reveals a characteristic behavior. Specifically, the critical groups of points are following: 2014 March 5.2 and 10.2, March 6.2 and 11.2, March 20.0 and 25.2, March 20.8-21.0 and 25.8-26.2, March 27.9-28.2 and April 2.0-2.1, March 28.9-29.1 and April 3.0-3.2, and April 29.2 and May 4.2. In these seven cases, the data taken at similar phases in the consecutive cycles differ by about 0.2 mag or more, which cannot be due to changes of aspect in only 5 days (one period) between the groups of points. This reveals that the asteroid was repeatedly seen not being in the same orientation at two times after one 122.2 h period. This is a characteristic lightcurve behavior of asteroid in a state of non-principal axis (NPA) rotation (see, e.g., Pravec et al. (2005, 2014) for more information on such asteroids in NPA rotation states, so called tumblers).

The lightcurve of 1999 NC43 resembles simulated lightcurves of tumblers in Short-Axis Mode (SAM) with the mean wobbling angle 20º -30º (Henych and Pravec, 2013). As discussed in Pravec et al. (2014), for the most prominent lightcurve period $P_1$ for a moderately excited SAM, it is $P_1^{-1} = P_\phi^{-1} - P_\psi^{-1}$, where $P_\phi$ is a period of rotation of the body around its shortest principal axis and $P_\psi$ is the time-averaged period of precession of this axis. We tentatively suggest that the NPA rotation of 1999 NC43 may be such a moderately excited SAM as well.

We attempted to derive the second period of the NPA rotation using the two-period Fourier series method (Pravec et al., 2005). We found a candidate $P_2$ of 200.3 h, but more data for the tumbler's lightcurve would be needed to establish this securely. Note that in the case of Apophis (Pravec et al., 2014), a much more thorough coverage obtained allowed us to derive a model of the tumbler, but such amount of data was practically impossible to obtain for the much longer period of 1999 NC43.

We derived the mean absolute magnitude of 1999 NC43 in 2000 and 2014 to be H = 16.24 ± 0.22 and 16.14 ± 0.21, respectively. The uncertainties are dominated by the assumed G uncertainty of ± 0.11 (see above). We obtained (V - R) = 0.448 ± 0.010, which is consistent with the Q classification of the asteroid.

The full observed amplitude of the lightcurve exceeds 1.2 mag. It indicates an elongated shape of the asteroid, but the amplitude was probably exaggerated by an amplitude-phase effect at the high solar phases of 40º to 80º. A correction of the amplitude to zero solar phase could be approximately done with the empirical formula by Zappalà et al. (1990). Assuming Zappalà et al.'s *m* parameter of 0.03, appropriate for SQ types, we obtained an estimate for the amplitude at zero solar phase of 0.5 - 0.6 mag. This corresponds to an axial ratio of the body of about 1.7.

*3.1.3 Implications of our Photometric Study*

The very slow rotation of 1999 NC43 could be a result of slow-down by the Yarkovsky-O'Keefe-Radzievskii-Paddack (YORP) effect. Regarding its NPA rotation state, we note that with its diameter and spin rate, the asteroid lies in the range where tumblers predominate (see Pravec et al., 2014), so it was no surprise that we found it tumbling. An estimated time of damping of its NPA rotation at its current spin rate is comparable to the age of the solar system (Pravec et al., 2014). This means that once its NPA rotation was set it could not be damped by the energy dissipation due to a stress-strain cycling within the tumbling body. However, a mechanism of the NPA rotation excitation is unclear. In Pravec et al. (2014), the following proposed excitation mechanisms were given: (1) original tumbling resulted from the formation of asteroid in disruption of its parent body, (2) sub-catastrophic impacts, (3) deceleration of the rotation rate by the YORP effect, and (4) gravitational torques during planetary flyby. The mechanism 1 does not seem plausible, as it is unlikely that the rotation of an asteroid fragment from catastrophic disruption would be nearly completely halted. The other three mechanisms appear possible. The current knowledge does not allow us to suggest which of them may be more likely for the case of 1999 NC43 and when it happened.

*3.2    Does the Q taxonomic type of 1999 NC43 mean it is an LL chondrite?*

An intriguing characteristic of 1999 NC43 is that this object has been classified as a Q-type (Binzel et al., 2004), a characteristic that would strengthen its possible link to the Chelyabinsk meteorite. Q-type asteroids are believed to have surface composition similar to ordinary chondrite meteorites (DeMeo et al., 2009). Ordinary chondrite meteorites include H, L and LL chondrites that have varying abundances and compositions of olivine and pyroxene, which is a function of the redox state during their formation. H chondrites are the most reduced with more metal, pyroxene and less olivine than L and LL chondrites. All three meteorite types are thought to have come from distinct source regions in the main asteroid belt. It has been argued that H chondrites come from the asteroid (6) Hebe (Gaffey and Gilbert, 1998) or from an asteroid family whose identity has yet to be determined (e.g., Rubin and Bottke, 2009); L chondrites come from the Gefion family (Nesvorný et al., 2009) and LL chondrites come from the Flora family (Bottke et al., 2002; Vernazza et al., 2008; de León et al., 2010; Dunn et al., 2013; Reddy et al. 2014).

While the Q-type taxonomic classification of 1999 NC43 (Binzel et al., 2004) is accurate, assuming that all Q-types are LL chondrites is not correct. To further explore this hypothesis we conducted an experiment where we taxonomically classified 38 ordinary chondrites (14-H, 14-L, and 10-LL chondrites) from Dunn et al. (2010) using the online Bus-DeMeo taxonomy calculator (http://smass.mit.edu/busdemeoclass.html). The Bus-DeMeo system was developed from the analysis of asteroid spectra (0.45-2.45 μm), and therefore, caution must be taken when using it with laboratory spectra of meteorite samples. However, our goal here is not to highlight the limitations of taxonomy but to stress the fact that assuming taxonomy is akin to surface composition could be misleading except in a few cases. The Bus-DeMeo online tool takes an uploaded spectrum and smoothes it using a cubic spline model. Then, the smoothed spectrum is normalized and the overall slope is removed so the principal component analysis can be applied. The taxonomic type assigned to each spectrum and the calculated principal components PC1' and PC2' are presented in Table 2.

Figure 4 shows a PC2' versus PC1' diagram from DeMeo et al. (2009), where the calculated PC values for H (red triangles), L (blue circles) and LL chondrites (black squares) are shown. Two key issues are revealed in this plot. First, the notion that all ordinary chondrites are Q-types is not correct; second not all LL chondrites are Q-types. This would open up the possibility for 1999 NC43 (while being a Q-type) to have a surface composition different from a LL5 chondrite like that of Chelyabinsk.

*3.3    Spectral and Compositional Affinity between Chelyabinsk and 1999 NC43*

*3.3.1    Telescopic Observations*

We attempted to obtain near-IR (0.7–2.5 µm) spectra of 1999 NC43 using the low-resolution SpeX instrument in prism mode (Rayner et al., 2003) on the NASA Infrared Telescope Facility (IRTF), Mauna Kea, Hawai'i between January 8-14, 2014. Unfortunately, the observing and weather conditions were not optimal for obtaining usable spectral data. However, we contacted the MIT-UH-IRTF Joint Campaign for NEO Reconnaissance and obtained the raw data for 1999 NC43. The NEA was observed by the campaign on March 23, 2007 using the NASA IRTF when it was 16.3 V magnitude. Apart from the asteroid, local standard and solar analog star observations were also performed to correct for telluric absorption features and solar continuum, respectively.

*3.3.2    Data Reduction and Analysis*

SpeX prism data were processed using the IDL-based Spextool provided by the NASA IRTF (Cushing et al., 2004). Analysis of the data to determine spectral band parameters like band centers, band depths and Band Area Ratio (BAR) was done using a Matlab code based on the protocols discussed by Cloutis et al. (1986). All-night average spectra were used to extract spectral band parameters. The calculated value for each band parameter corresponds to an average value estimated from ten measurements obtained using different order polynomial fits (typically third and fourth order), and sampling different ranges of points within the corresponding intervals. The errors for the band parameters were determined from the multiple measurements and are given by the standard deviation of the mean (1-σ). We estimated 1-σ errors of 0.01 µm for the Band I center and 0.04 for the BAR given the point-to-point scatter in the data. Figure 5 shows the near-IR spectrum of 1999 NC43 with two absorption bands due to the minerals olivine and pyroxene.

Spectral band parameters (Table 3) are consistent with those from de León et al. (2010) as they used the same data set for their analysis. The only parameter that differed between the two is BAR. This is because de León et al. (2010) used Band II area definition from Cloutis et al. (1986) where the area of Band II was calculated using a straight line continuum up to 2.42 µm. In contrast, we used the full band till 2.5 µm to estimate our Band II area consistent with the methods used by Dunn et al. (2010). Hence, the BAR value estimated by de León et al. (2010) is lower than what we estimated using the newer definition, although the conclusions drawn from both are identical.

Temperature-induced spectral effects can cause changes in spectral band parameters (e.g., Reddy et al., 2012c; Sanchez et al., 2012). We used equations from Burbine et al. (2009) to calculate the surface temperature of 1999 NC43 at the time of observations and equations from Sanchez et al. (2012) to correct for temperature effects and normalize the data to room temperature before analysis using laboratory spectra. Temperature effect on Band I center is negligible (0.0003 µm) compared to the wavelength precision of laboratory or observational data. We applied temperature corrections for Band II center and BAR using Sanchez et al. (2012) equations. Since the olivine and pyroxene chemistries are calculated using Band I center, the effect of temperature on our actual analysis is negligible.

### 3.3.3 Laboratory Spectra: Methodology

Reflectance spectra of Chelyabinsk meteorite samples were acquired with an Analytical Spectral Devices FieldSpec Pro HR spectrometer over the range of 350 to 2500 nm in 1.4 nm steps, with a spectral resolution of 2 to 7 nm. The data are internally resampled by the instrument to output data at 1 nm intervals. All spectra were measured at a viewing geometry of $i = 30°$ and $e = 0°$ where the incident lighting was provided by an in-house 100W quartz–tungsten–halogen collimated light source. Sample spectra were measured relative to a 100% Labsphere Spectralon® standard and corrected for the minor (<2%) irregularities in its absolute reflectance and occasional detector offsets at 1000 and 1830 nm. In each case, 500 spectra of the dark current, standard, and sample were acquired and averaged, to provide sufficient signal-to-noise. All mixtures had grain size of <45µm. Spectra of the two dominant lithologies of Chelyabinsk meteorite, an LL5 chondrite lithology and a dark, fine-grained 'impact melt' lithology (containing both bona fide impact melt and impact-affected 'shock blackened' material), as well as known-proportion mixtures of the two lithologies were collected. Figure 6 shows the end member spectra of LL5 chondrite matrix and impact melt from Chelyabinsk. The spectra have been normalized to unity at 1.5 µm.

### 3.3.4 Does 1999 NC43 have the same spectral properties as Chelyabinsk?

Curve matching is a non-diagnostic tool that can provide general clues to the possible nature of the asteroid's surface when there are no strong diagnostic mineral absorption features (Gaffey, 2008). The spectrum of 1999 NC3 (Figure 5) shows strong absorption bands at 1 µm (Band I) and 2 µm (Band II) due to the minerals olivine and pyroxene (Gaffey, 1984). The depth of an absorption feature is a non-diagnostic parameter that is influenced by a host of factors including: a) abundance of the absorbing species (Reddy et al., 2010); b) particle size (Reddy et al., 2012a); c) phase angle (Reddy et al., 2012c; Sanchez et al., 2012); d) abundance of opaques such as carbon and metal (Reddy et al., 2012b; Le Corre et al., 2011); and impact shock and melt (Britt and Pieters, 1989). We fitted a continuum-removed spectrum of laboratory mixtures of 70% Chelyabinsk LL5 chondrite and 30% impact melt to the continuum-removed 1999 NC43 spectrum (Figure 7). We picked this mixture because it has been generally observed that this is the ratio between shocked/impact melted and unaltered LL5 chondrite material (Kohout et al., 2013). So a 70:30 mixture represents a typical whole rock spectrum. Chelyabinsk Band I depth (14%) is shallower than that from 1999 NC43 (25%), suggesting that the asteroid might have little or no impact melt in its surface assemblage.

To test this hypothesis we compared the spectra of 1999 NC43 with that of 100% LL5 chondrite matrix from Chelyabinsk (see Fig. 8). Similar to the previous mixture, the 100% LL5 spectrum has a Band I depth (19%) that is shallower than that of 1999 NC43 (25%). This confirms that there is little or no impact melt on the surface of 1999 NC43. We then looked at a range of particle sizes for 100% LL5 chondritic spectrum from Chelyabinsk to determine if particle size could cause the mismatch. The closest match to the asteroid Band I depth (25%) is the spectrum of a slab with a Band I depth of 22%. However, the Band II is nonexistent in this slab spectrum, whereas it is 18% deep in 1999 NC43. This rules out particle size as a cause for the mismatch between the spectra of Chelyabinsk and 1999 NC43. The lack of a spectral match between Chelyabinsk and 1999 NC43 suggests that the asteroid might not be the source object for this meteorite. To further verify this, we conducted a detailed mineralogical analysis of the two spectra and confirmed the lack of affinity between the two, as described in the next section.

### 3.3.5  Do 1999 NC43 and Chelyabinsk have the same composition?

#### 3.3.5.1 Surface Composition

Surface mineralogy is the most robust way of identifying meteorite analogs of asteroids using remote sensing. Spectral band parameters are the first step towards identification of meteorite analogs for asteroids. Gaffey et al. (1993) classified S-type asteroids into subtypes based on their Band I center and BAR values. Ordinary chondrite like asteroids fall into the S(IV) category with Band I center increasing and BAR decreasing as we go from H, to L, to LL chondrites due to increasing olivine abundance. Dunn et al. (2010) plotted ordinary chondrites in this parameter space to show the zones within the S(IV) region for H, L and LL chondrites (Fig. 9). We have adapted this figure and plotted the values for Chelyabinsk (whole rock) and 1999 NC43. The asteroid parameters are located at the edge of the L and H chondrite zone, whereas Chelyabinsk is located in the LL chondrite zone, consistent with its laboratory classification.

We used Dunn et al. (2010) equations to derive the chemistry of olivine and pyroxene from spectra of ordinary chondrites. These equations have been successfully applied to derive surface composition of several asteroids (e.g., Reddy et al., 2011; Sanchez et al., 2013; Dunn et al. 2013; Reddy et al. 2014). Using the Band I center, olivine (fayalite) and pyroxene (ferrosilite) mol. % can be estimated with uncertainties of ±1.3 and ±1.4 mol. %, respectively. Reddy et al. (2014) verified the validity of these equations by comparing spectrally derived Fa and Fs values from Binzel et al. (2001) ($Fa_{28.40±1.3}$ and $Fs_{23.35±1.4}$) of near-Earth asteroid (25143) Itokawa, to those measured from samples of Itokawa ($Fa_{28.60}$ and $Fs_{23.10}$) returned from the Hayabusa spacecraft (Nakamura et al. 2011). The difference between laboratory measured values of Itokawa samples and those from IRTF spectral data is < 1 mol. % confirming Itokawa as an LL chondrite and attesting to the validity of the method we applied to constrain surface mineralogy.

Using the same technique we extracted the spectral band parameters for Chelyabinsk, (8) Flora, and 1999 NC43 and calculated the (Fa) olivine and (Fs) pyroxene chemistry (Fig. 10). While, Chelyabinsk (spectrally derived $Fa_{30.6}$ and $Fs_{25.0}$) and (8) Flora ($Fa_{29±1.3}$ and $Fs_{24±1.4}$) plot in the LL chondrite zone as expected, 1999 NC43 ($Fa_{23.2±1.3}$ and $Fs_{19.6±1.4}$) is located in the L chondrite zone away from Chelyabinsk. This strongly suggests that the surface composition of 1999 NC43 is similar to L chondrites rather than LL chondrites such as Chelyabinsk (Fig. 10).

Another parameter that could be useful for distinguishing H, L and LL chondrites is the olivine abundance. Dunn et al. (2010) developed equations to estimate the olivine abundance in ordinary chondrites using their Band Area Ratio. We used this parameter to further verify that Chelyabinsk and 1999 NC43 are compositional related. Figures 11 and 12 are plots of olivine and pyroxene chemistries as a function of olivine abundance in ordinary chondrites. These two figures have been adapted from Dunn et al. (2010) and show the zones for H, L, and LL chondrites. The gray boxes indicate the uncertainties associated with these parameters. In Fig. 11, Chelyabinsk is located in the LL chondrite zone whereas 1999 NC43 plots close to the border between L and H chondrite region. The situation is similar when we plot pyroxene chemistry vs. olivine abundance in Fig. 12 where the NEA is located far from the LL chondrite region, away from Chelyabinsk. These two plots further indicate that 1999 NC43 is unlikely to be the source asteroid for the Chelyabinsk meteoroid based on the compositional differences between the two.

*3.3.5.2 Modified Gaussian Model Analysis*

The Modified Gaussian Model (MGM) was developed to allow reflectance spectra to be deconvolved into component absorption bands that could be directly attributed to specific electronic transitions (Sunshine et al., 1990). The reflectance spectrum is converted to natural log reflectance to approximate absorption. To fit a spectrum, the user defines the starting centers, widths, and strengths of each modified Gaussian absorption band, the slope and offset of a continuum line, and uncertainty values within which each of the parameters will be varied as the model searches for the best fit to the spectrum. All values are adjusted until the residual error between the fit and the spectrum is minimized, or until the difference between subsequent solutions is less than a user-defined threshold value.

Because the MGM sums the absorption bands and continuum to create a mathematically accurate spectral fit, solutions either need to be restricted to the minimum number of bands or incorporate additional constraints imposed by crystal chemistry of the minerals present for the fits to be physically useful. To assess the overall proportions of olivine to pyroxene in 1999 NC43, we first conducted a simplified MGM fit using five bands, similar to the approach used by Binzel et al. (2001). One band was placed in the visible to account for intervalence charge transfers and residual continuum curvature, three were placed in the 1 μm region to account for olivine and pyroxene absorptions, and one was placed in the 2 μm region to account for pyroxene absorptions. Though the three absorption bands near 1 μm are the result of three olivine absorptions and at least two pyroxene absorptions that occur at very similar energies, the longest wavelength band, near 1.2 μm, is primarily due to olivine absorptions and the shortest wavelength band, near 0.9 μm, is primarily due to orthopyroxene. Thus, the relative proportions of each mineral can be inferred by the relative strengths of these bands (Binzel et al., 2001). Using the band intensities of the absorption bands at 0.9 μm and 1.2 μm, we calculate an $s_{olv}/s_{pyx}$ to be 0.8, which, based on relationships for H, L, and LL chondrites determined by Binzel et al. (2001), suggests that 1999 NC43 is compositionally intermediate to the L and LL chondrite averages.

We have also attempted a more complicated eight band fit in which bands associated with each component mineral were constrained to be positioned at physically realistic energies for each mineral group (Fig. 13). The band centers and allowable variation for the modified Gaussians associated with olivine were restricted to those measured for natural (Sunshine and Pieters, 1993) and synthetic (Isaacson et al., 2014) olivines, and those associate with pyroxenes were restricted to be consistent with those measured for synthetic, pure quadrilateral pyroxenes by Klima et al. (2007; 2011). Though the error ranges were restricted, the model is not currently designed to force parameter covariance. It is important to note that despite the addition of two bands in the 1-μm region, the overall fit is not significantly improved relative to the simplified fit. Taken at face value, the parameters derived for the olivine and pyroxene components of the eight-band fit suggest a relatively high proportion of olivine to pyroxene and a relatively high abundance of iron in the orthopyroxene (~$En_{70}$). Because the wavelength positions of each component of the 1 μm olivine absorption band varies by only about 100 nm at most over the full range of olivine compositions, a small uncertainty in band center translates to a large uncertainty in Fo content of the olivine. However, based on these positions, the olivine composition is interpreted as $Fo_{75\pm15}$. These values suggest that 1999 NC43 may lie within the LL chondrite compositional range, though the L chondrite range is also within the uncertainty. It is important to note that MGM does not account for changes in a spectrum due to non-compositional effects such as temperature and phase angle which are know to change the band width, band depth and spectral slope (Sanchez et al. 2012; Reddy et al. 2012c). Our derived ambiguity between L and LL chondrite options could be partly explained if one would invoke these effects. Mineralogy derived from spectral band parameters corrected for these effects is more robust.

## 4. Discussion

### *4.1 Rotational Spectral Variation*

Our compositional analysis suggests that 1999 NC43 does not have a surface composition similar to an LL chondrite. However, there is a non-zero probability for rotational spectral variation involving L and LL chondrite material. Rotational spectral variation is typically interpreted as a change in surface composition with rotation phase. Compositional variation is primarily due to variation in diagnostic spectral band parameters as the asteroid rotates. These diagnostic spectral band parameters are Band I and II centers for olivine and pyroxenes and Band Area Ratio for olivine/pyroxene mixtures. Variations in non-diagnostic spectral parameters such as spectral slope could be due to non-compositional effects.

The only main belt asteroid with confirmed rotational compositional variation is (4) Vesta (Gaffey 1997; Binzel et al. 1997; Reddy et al. 2010b; Li et al. 2010). This was first observed in ground based rotational spectra of the asteroid and later confirmed by the Dawn mission (Reddy et al. 2012a; De Sanctis et al. 2012). In addition to Vesta, spacecraft observations of asteroid (21) Lutetia showed variations in albedo of up to 30% (Sierks et al. 2011), however from the available data it was not possible to confirm whether these variations were caused by differences in composition, regolith grain size, or due to space weathering (Barucci et al. 2012). Among the NEA population, asteroid 2008 TC3 is the only confirmed case of compositional variations. This asteroid was discovered just a few hours before it impacted the Earth on October 7th 2008. Fragments recovered from the impact site (now known as the Almahata Sitta meteorite) were classified as polymict ureilites, and their analysis showed that this meteorite is composed of several different achondritic and chondritic lithologies (Bischoff et al. 2010). Reflectance spectra of this object indicated that it belongs to the uncommon class of F-type asteroids (Jenniskens et al. 2009). Most near-Earth asteroids are much smaller than Vesta and Lutetia, with diameters typically <2 km (Binzel et al. 2002), and a large fraction of these objects belong to the rather common S-complex. These objects are thought to be fragments from multiple collisions of their parent body in the asteroid belt (Binzel et al. 2002, Davies et al., 2002). These multiple collisions would lead to the formation of small bodies with a likely rubble pile structure, in which all the material has been mixed (Leinhardt et al. 2000, Richardson et al. 2002). Most objects that become NEAs develop highly eccentric orbits with aphelia in the main belt. As a result, these NEAs are expected to experience further collisions, with collision velocities much higher than those among main belt objects (Bottke et al., 1994; Dell'Oro et al., 2011).

Based on numerical simulations Leinhardt et al. (2000) showed that collisions between rubble pile objects at high velocities will produce larger disruptions and increased mixing of the material. Thus, when we take into account the collisional nature of NEA formation (and subsequent evolution), hemispherical compositional variations are unlikely to be seen. This was the case for two different NEAs, (433) Eros and (25143) Itokawa, that were visited by spacecraft. NIR spectroscopic observations of (433) Eros obtained by the NEAR Shoemaker mission found no evidence of compositional variations on the surface of this asteroid. Regarding these results, Bell et al. (2002) stated: "*overall, the surface of Eros appears remarkably spectrally homogeneous*". For asteroid Itokawa, albedo and color variations were observed by the Hayabusa spacecraft. However, these variations were not attributed to bulk compositional heterogeneity, but resulted from differences in the physical properties and microstructure of the material, i.e., differences in grain sizes and/or the degree of space weathering (Abe et al. 2006). These variations were too subtle to be detected from ground-based telescopic observations (Binzel et al. 2001; Abe et al. 2006).

Evidence for potential rotational spectral variations could come in the form of exogenic clasts in meteorites. A survey of 16983 L chondrite meteorites shows that LL chondrite xenoliths are extremely rare, with only three possible L chondrite breccias (Paragould, Adzhi-Bogdo and NWA 6728) with LL chondrite xenoliths. If there were significant collisional mixing of L and LL chondrite material then one would expect to see abundant evidence for this in the meteorites delivered to the Earth. The rarity of L/LL collisional mixtures suggests that rotational spectral variation involving these two meteorite types on 1999 NC43 is unlikely. A final piece of evidence comes from Chelyabinsk meteorite itself where there is no evidence for the presence of L chondrite material. In any case, the slow rotation of 1999 NC43, reported in Sec. 3.1.2, makes future observational efforts to probe possible rotational spectral variations rather easy.

We also rule out space weathering effects as a way to explain the spectral differences between 1999 NC43 and Chelyabinsk. The observed spectral differences between the two are in diagnostic spectral band parameters such as band centers. Gaffey (2010) has shown that space-weathering effects do not influence spectral band parameters. So even if 1999 NC43's surface was space weathered it won't affect the spectral band parameters to make it an L chondrite instead of an LL chondrite like Chelyabinsk.

*4.2    Compositional Variability Among Asteroid Pairs*

Another less likely possibility is that Chelyabinsk is an LL chondrite fragment that was ejected from an L chondrite type asteroid (1999 NC43). Asteroid pairs are two objects with osculating orbital elements too similar to be due to random fluctuations of background asteroid population density (Vokrouhlický and Nesvorný, 2008). Compositional similarities between the two objects has been a powerful tool to verify their common origin and several studies have focused on this effort (e.g., Moskovitz et al. 2012; Duddy et al. 2013; Polishook et al. 2013; Wolters et al. 2014). Moskovitz et al. (2012) observed eleven asteroid pairs and found that 98% of them had similar BVRI colors. Wolters et al. (2014) obtained near-IR spectra of several asteroid pairs and suggested that some asteroid pairs showed differences in composition based on taxonomic classification and curve matching. Though no detailed mineralogical analysis was performed on these asteroid pairs, the spectral difference between the primary and secondary of some asteroid pairs indicates the presence of some surface heterogeneity. While the observed spectral difference between components of some asteroid pairs leaves open the possibility of a common origin, more rigorous mineralogical analysis of asteroid pairs is needed to evaluate the plausibility of 1999 NC43 and the Chelyabinsk meteoroid as an asteroid pair.

*4.3   Cosmic Ray Exposure Ages*

Another constraint on the origin of the Chelyabinsk bolide comes from the cosmic ray exposure (CRE) ages of the recovered samples. The penetration depth of cosmic rays into rock is on the order of a meter or so, so thus CRE ages represent a measure of cosmic ray damage to the body in question. When free-floating bodies in space are a few meters in diameter, cosmic ray damage comes from isotropic directions, and is referred to as 4 pi exposure. When the body is larger, such that only the surface experiences cosmic ray hits from more limited directions, the damage is called 2 pi exposure. According to Popova et al. (2013), several Chelyabinsk fragments show CRE ages of 1.2 My from 2 pi exposure. Other samples have limited or no exposure, implying they resided in the interior of a body that was at least 6 m is diameter.

The CRE ages of Chelyabinsk materials appear incompatible with the very short separation timescales discussed in Borovička et al. (2013) between the 2 km diameter asteroid 1999 NC43 and the 20 m diameter Chelyabinsk bolide. We assert that the vast majority of surface material on a Chelyabinsk bolide ejected by a cratering event could only begin receiving 2 pi exposure once it was separated from its immediate precursor. This implies a separation time of 1.2 My, too long for the orbits of the aforementioned bodies to remain compatible with one another, according to the orbital integration work presented here.

**5.   Summary**

Our statistical, dynamical and compositional study indicates that the suggested link between Chelyabinsk body and the asteroid 1999 NC43 is unlikely. While their orbits indeed prove to be at an anomalously small distance from each other, with a probability $\sim(1-3) \times 10^{-4}$ to occur by chance, likelihood of the implied hypothesis of Chelyabinsk being an ejected fragment from 1999 NC43 is nevertheless small. Our calculation yields an estimated probability of $\sim 10^{-3}$ to $10^{-2}$ for an origin of the Chelyabinsk impactor from 1999 NC43. Obviously, arguments based on probability estimates have their limitations, and to proceed further in assessing the proposed link requires further analyses.

For example, it is important to assess the likelihood of past orbital convergence of the Chelyabinsk pre-atmospheric orbit and that of 1999 NC43. Borovička et al. (2013) observed that the orbital similarity between the bodies persisted over 2,000 years, and that their distance allowed ejection velocity of the Chelyabinsk body from 1999 NC43 between $\simeq 0.7$ km/s and $\simeq 2$ km/s. However, we believe such a criterion is too weak, because the velocity threshold used is too high for typical collisional ejecta. In the putative case of a body in the size range of 1999 NC43, this would imply separation velocities between 1-2 m/s, or smaller. It is very difficult to reach this level of convergence, which shows the difficulties in convincingly demonstrating a dynamical link between two bodies in NEO space in general.

Another pathway for approaching the problem is to extend the investigation of compositional similarity of the two bodies beyond what we have done. We note that the Q-type taxonomic classification of 1999 NC43 (Binzel et al. 2004) is accurate. However, assuming that all Q-types are LL chondrites is not accurate. Taxonomic classification of the reflectance spectra of 38 ordinary chondrites (14-H; 14-L, and 10-LL chondrites) demonstrates that a) not all ordinary chondrites are Q-types; and b) not all LL chondrites are Q-types. This suggests that 1999 NC43 (while being a Q-type), can have a surface composition different from an LL5 chondrite like that of Chelyabinsk. Spectral curve matching between the continuum-removed spectra of any combination of Chelyabinsk LL5 chondrite and its impact melt to the spectrum of 1999 NC43 fails to produce a match. Similar curve matching of spectra of different particle sizes of Chelyabinsk meteorite also failed to yield a positive match. Finally, mineralogical analysis of Chelyabinsk and (8) Flora (the largest member of the presumed parent family) shows that their olivine and pyroxene chemistries are similar to LL chondrites. However, similar analysis of 1999 NC43 shows that its olivine and pyroxene chemistries are similar to L chondrites rather than an LL chondrite like Chelyabinsk. While MGM analysis suggests 1999 NC43 is similar to LL or L chondrite, we suspect this ambiguity is due to lack of temperature and phase angle corrections in the model.

While we acknowledge that more studies are needed, the lack of spectral match between Chelyabinsk and 1999 NC43, and their different chemistries described in this paper suggest that the asteroid is probably not the source object for this meteorite.

## Acknowledgements

This research work was supported by NASA Planetary Mission Data Analysis Program Grant NNX13AP27G, NASA NEOO Program Grant NNX12AG12G, NASA Planetary Geology and Geophysics Grant NNX11AN84G, the Grant Agency of the Czech Republic (grants P209-12-0229 and P209-13-01308S), and by the Ministry of Education of the Czech Republic, Grant LG12001. We thank the IRTF TAC for awarding time to this project, and to the IRTF TOs and MKSS staff for their support. Part of the data utilized in this publication were obtained and made available by the MIT-UH-IRTF Joint Campaign for NEO Reconnaissance. The IRTF is operated by the University of Hawaii under Cooperative Agreement no. NCC 5-538 with the National Aeronautics and Space Administration, Office of Space Science, Planetary Astronomy Program. The MIT component of this work is supported by NASA grant 09-NEOO009-0001, and by the National Science Foundation under Grants Nos. 0506716 and 0907766. EAC wishes to thank CFI, MRIF, CSA, and the University of Winnipeg for their support in establishing the University of Winnipeg Planetary Spectrophotometer Facility, and CSA and NSERC for supporting this study. Finally, we thank reviewer Jiří Borovička, and coauthors of his *Nature* paper, Pierre Vernazza, Brian Burt, Francesca DeMeo, Rick Binzel and Hal Levison for numerous discussions and especially for pointing out an error in the first version of this paper. We also thank Tom Kaye for help with the observations of 1999 NC43.

**Figure Captions**

Figure 1. The $\chi^2$ vs. period of the 2$^{nd}$-order Fourier series fitted to the reduced 2014 data for (86039) 1999 NC43. A period of 122.2 h and its half provide the best fit, but they do not explain the data fully (see text).

Figure 2. The reduced R data from 2014 folded with the period of 122.2 h. The curve is the best fit 2nd-order Fourier series. The character of the scatter of points around the mean curve reveals that the asteroid is tumbling.

Figure 3. This figure is same as Fig. 2, but for the data from 2000.

Figure 4. Plot showing the principal components (PC2' versus PC1' diagram) after DeMeo et al. (2009), where the calculated PC values for LL chondrites are depicted as black squares; L chondrites as blue circles and H chondrites as red triangles. Not all ordinary chondrites fall in the field of Q-types.

Figure 5. Near-IR spectrum of (86039) 1999 NC43 obtained using the SpeX instrument from the NASA IRTF. The spectrum shows two absorption bands near 1 and 2 µm due to the presence of the minerals olivine and pyroxene. The data were obtained and made available by the MIT-UH-IRTF Joint Campaign for NEO Reconnaissance.

Figure 6. End-member spectra of LL5 chondrite and impact melt lithologies from Chelyabinsk normalized to unity at 1.5 µm. The spectra show two absorption bands near 1 and 2 µm due to the presence of the minerals olivine and pyroxene.

Figure 7. Continuum-removed spectrum of (86039) 1999 NC43 overlaid with a continuum-removed laboratory spectrum (<45 µm) of a 70:30 mixture of Chelyabinsk LL5 chondrite and impact melt. The Chelyabinsk meteorite contains approximately 1/3rd impact melt and 2/3rd LL5 chondrite lithologies. Band depths of the Chelyabinsk meteorite do not match with those of 1999 NC43 suggesting a different meteorite type/mafic mineral abundance/particle size/impact melt abundance.

Figure 8. Continuum removed spectrum of (86039) 1999 NC43 overlaid with a continuum-removed laboratory spectrum of <45 µm size 100% Chelyabinsk LL5 chondrite. Even after eliminating the impact melt from the mixture, the band depths of 100% Chelyabinsk LL5 chondrite are too shallow compared to those of 1999 NC43 suggesting that impact melt abundance is not the cause of the band depth mismatch.

Figure 9. Band I center vs. Band Area Ratio plot showing the Gaffey S(IV) region with data for H, L and LL chondrites from Dunn et al. (2010). The dashed line indicates the location of the olivine-orthopyroxene mixing line of Cloutis et al. (1986). Chelyabinsk band parameters plot in the LL chondrite zone consistent with its classification. (86039) 1999 NC43 plots on the edge of the L chondrite zone suggesting that its surface composition is inconsistent with LL chondrite and making it unlikely to be parent asteroid of Chelyabinsk.

Figure 10. Plot showing olivine iron content (mol.% fayalite) on the Y-axis and pyroxene iron content (mol.% ferrosilite) on the X-axis from laboratory measurements of ordinary chondrites (H, L and LL). Fayalite and ferrosilite values for Chelyabinsk (spectrally derived from whole rock sample), asteroid (8) Flora (the source of the LL chondrites), and (86039) 1999 NC43 are shown. The error bars in the upper right corner correspond to the uncertainties derived by Dunn et al. (2010), 1.3 mol% for Fa, and 1.4 mol% for Fs. The spectrally derived Fs and Fa values for Chelyabinsk fall in the LL chondrite zone along with (8) Flora. In contrast, the Fs and Fa values of (86039) 1999 NC43 fall in the L chondrite zone away from Chelyabinsk or (8) Flora, suggesting that it is unlikely to be the parent asteroid of the Chelyabinsk meteoroid. Figure adapted from Nakamura et al. (2011).

Figure 11. Spectrally-derived olivine chemistry (Fa) vs. olivine abundance plot for ordinary chondrites showing Chelyabinsk and (86039) 1999 NC43. Dashed boxes represent the approximated range of measured values for ordinary chondrites. The gray boxes indicate the uncertainties associated to the spectrally-derived values. Asteroid 1999 NC43 plots near the border between L and H chondrites, whereas Chelyabinsk plots clearly in the heart of the LL chondrite zone. This figure has been adapted from Dunn et al. (2010).

Figure 12. Spectrally-derived pyroxene chemistry (Fs) vs. olivine abundance plot for ordinary chondrites showing Chelyabinsk and (86039) 1999 NC43. Dashed boxes represent the approximated range of measured values for ordinary chondrites. The gray boxes indicate the uncertainties associated to the spectrally-derived values. Asteroid 1999 NC43 plots near the border between L and H chondrites, whereas Chelyabinsk plots clearly in the heart of the LL chondrite zone. This figure has been adapted from Dunn et al. (2010).

Figure 13. MGM fits for 1999 NC43 showing the measured spectrum is shown in gray circles, the continuum is shown as a dotted black line, and the modeled fit is shown as a solid black line. The RMS error as a function of wavelength is offset up by 0.15 log reflectance units and is shown as a thin black line. Absorptions attributed to olivine are depicted in green, orthopyroxene in red, and clinopyroxene in blue. The extensive overlap of the bands leads to large uncertainties in the exact positioning of the component absorption bands near 1 μm.

**Tables**

Table 1. Observational circumstances for photometric data

| Observation Date. | Filter | Observation Location | Telescope Size |
|---|---|---|---|
| 2000-02-27.8 | R | Ondřejov | 0.65-m |
| 2000-02-28.8 | R | Ondřejov | 0.65-m |
| 2000-03-02.8 | R | Ondřejov | 0.65-m |
| 2000-03-04.8 | R | Ondřejov | 0.65-m |
| 2000-03-06.8 | R | Ondřejov | 0.65-m |
| 2000-03-10.8 | R | Ondřejov | 0.65-m |
| 2000-03-22.8 | R | Ondřejov | 0.65-m |
| 2000-03-23.8 | R | Ondřejov | 0.65-m |
| 2000-04-02.9 | R | Ondřejov | 0.65-m |
| 2000-04-06.8 | R | Ondřejov | 0.65-m |
| 2000-04-08.8 | R | Ondřejov | 0.65-m |
| 2014-03-03.1 | r' | Hereford | 0.30-m |
| 2014-03-04.1 | r' | Hereford | 0.30-m |
| 2014-03-05.2 | r' | Hereford | 0.30-m |
| 2014-03-06.2 | r' | Hereford | 0.30-m |
| 2014-03-07.2 | r' | Hereford | 0.30-m |
| 2014-03-08.1 | r' | Hereford | 0.30-m |
| 2014-03-09.3 | r' | Hereford | 0.30-m |
| 2014-03-10.2 | r' | Hereford | 0.30-m |
| 2014-03-11.2 | r' | Hereford | 0.30-m |
| 2014-03-15.2 | r' | Lowell | 0.70-m |
| 2014-03-16.2 | r' | Lowell | 0.70-m |
| 2014-03-17.2 | r' | Lowell | 0.70-m |
| 2014-03-18.2 | r' | Lowell | 0.70-m |
| 2014-03-19.2 | r' | Lowell | 0.70-m |
| 2014-03-19.9 | R | Ondřejov | 0.65-m |
| 2014-03-20.0 | R | La Silla | 1.54-m |
| 2014-03-20.8 | R | Ondřejov | 0.65-m |
| 2014-03-21.0 | R | La Silla | 1.54-m |
| 2014-03-25.2 | r' | Lowell | 0.70-m |
| 2014-03-25.8 | R | Ondřejov | 0.65-m |
| 2014-03-26.1 | R | La Silla | 1.54-m |
| 2014-03-26.2 | r' | Lowell | 0.70-m |
| 2014-03-27.9 | R | Ondřejov | 0.65-m |
| 2014-03-28.1 | R | La Silla | 1.54-m |
| 2014-03-28.9 | R | Ondřejov | 0.65-m |
| 2014-03-29.1 | R | La Silla | 1.54-m |
| 2014-03-29.8 | R | Ondřejov | 0.65-m |
| 2014-03-30.1 | R | La Silla | 1.54-m |
| 2014-03-31.1 | R | La Silla | 1.54-m |
| 2014-04-01.0 | R | La Silla | 1.54-m |
| 2014-04-02.1 | R | La Silla | 1.54-m |
| 2014-04-03.1 | R | La Silla | 1.54-m |
| 2014-04-04.0 | R, V | La Silla | 1.54-m |
| 2014-04-28.2 | r' | Hereford | 0.30-m |
| 2014-04-28.6 | r' | Australia | 0.30-m |
| 2014-04-29.2 | r' | Hereford | 0.30-m |
| 2014-04-29.5 | r' | Australia | 0.30-m |
| 2014-04-30.2 | r' | Hereford | 0.30-m |
| 2014-05-01.2 | r' | Hereford | 0.30-m |
| 2014-05-01.6 | r' | Australia | 0.30-m |
| 2014-05-02.2 | r' | Hereford | 0.30-m |
| 2014-05-03.2 | r' | Hereford | 0.30-m |
| 2014-05-04.2 | r' | Hereford | 0.30-m |
| 2014-04-30.2 | r' | Hereford | 0.30-m |
| 2014-05-01.2 | r' | Hereford | 0.30-m |
| 2014-05-01.6 | r' | Australia | 0.30-m |

Table 2. Taxonomic classification of ordinary chondrite meteorites. Names and details of samples are published in Dunn et al. (2010).

| Type | RELAB ID | Taxonomy | PC1' | PC2' |
|---|---|---|---|---|
| LL4 | MT-HYM-075 | S-complex, Sq-type | −0.2412 | 0.1009 |
| LL4 | MT-HYM-083 | S-complex, Sq-type | −0.3935 | 0.1660 |
| LL4 | TB-TJM-075 | Q-type | −0.5977 | 0.3303 |
| LL5 | MT-HYM-077 | S-complex, Sq-type | −0.2313 | 0.0908 |
| LL5 | MT-HYM-079 | Q-type | −0.6220 | 0.2023 |
| LL5 | MT-HYM-080 | Q-type | −0.8158 | 0.3465 |
| LL5 | MT-HYM-085 | Q-type | −0.7781 | 0.3407 |
| LL6 | TB-TJM-145 | Q-type | −0.7334 | 0.2161 |
| LL6 | TB-TJM-067 | Q-type | −0.7126 | 0.2510 |
| LL6 | TB-TJM-077 | Q-type | −0.7948 | 0.2499 |
| L4 | TB-TJM-102 | S-complex, Sq-type | −0.3060 | 0.0745 |
| L4 | TB-TJM-121 | S-complex, S-type | −0.0456 | −0.0651 |
| L4 | TB-TJM-065 | S-complex, Sr-type | −0.1885 | 0.2461 |
| L4 | TB-TJM-081 | S-complex, Sr-type | −0.0695 | 0.1621 |
| L5 | MT-HYM-084 | S-complex, Sr-type | −0.3740 | 0.3347 |
| L5 | TB-TJM-109 | S-complex, Sq-type | −0.3760 | 0.1942 |
| L5 | TB-TJM-134 | S-complex, Sr-type | −0.3234 | 0.2681 |
| L5 | TB-TJM-099 | Q-type | −0.4467 | 0.1942 |
| L6 | TB-TJM-101 | Q-type | −0.6916 | 0.4465 |
| L6 | TB-TJM-130 | S-complex, Sr-type | −0.2586 | 0.3275 |
| L6 | TB-TJM-137 | S-complex, Sq-type | −0.1835 | 0.0039 |
| L6 | TB-TJM-139 | Q-type | −0.6346 | 0.4029 |
| L6 | TB-TJM-140 | S-complex, Sr-type | −0.0687 | 0.2857 |
| L6 | TB-TJM-064 | S-complex, Sr-type | −0.1736 | 0.2807 |
| H4 | TB-TJM-128 | S-complex, Sq-type | −0.2734 | 0.1363 |
| H4 | TB-TJM-136 | Q-type | −0.4609 | 0.1734 |
| H4 | TB-TJM-082 | S-complex, S-type | 0.4020 | −0.0419 |
| H4 | TB-TJM-083 | S-complex, Sv-type | 0.7910 | −0.0328 |
| H5 | TB-TJM-104 | Q-type | −0.4104 | 0.3266 |
| H5 | TB-TJM-129 | S-complex, Sq-type | −0.2040 | 0.1331 |
| H5 | TB-TJM-143 | S-complex, S-type | −0.0349 | 0.1120 |
| H5 | TB-TJM-074 | O-type or Q-type | −0.6051 | 0.4531 |
| H5 | TB-TJM-097 | S-complex, Sr-type | 0.4466 | −0.0185 |
| H6 | TB-TJM-131 | O-type or Q-type | −0.7851 | 0.7458 |
| H6 | TB-TJM-135 | S-complex, S-type | 0.2463 | 0.0213 |
| H6 | TB-TJM-069 | S-complex, S-type | 0.2887 | −0.0847 |
| H6 | TB-TJM-088 | S-complex, Sq-type | −0.3284 | 0.1805 |
| H6 | TB-TJM-094 | O-type or Q-type | −0.8736 | 0.6176 |

Table 3. Spectral band parameters and mineralogy

| Asteroid/Sample | Band I Center | Band II Center | BAR* | Olivine/(Olv+Pyx) | Fayalite | Ferrosilite |
|---|---|---|---|---|---|---|
| Units | μm | μm | | | Mol.% | Mol.% |
| (25143) Itokawa (Binzel et al. 2001) | 0.99±0.01 | 2.02±0.03 | 0.39±0.04 | 0.63±0.03 | 28.4±1.3 | 23.3±1.4 |
| Itokawa Samples (Nakamura et al.) | | | | | 28.6±1.1 | 23.1±2.2 |
| Chelyabinsk (Spec. Derived from whole rock sample) | 1.02±0.01 | 1.90±0.03 | 0.43±0.05 | 0.62±0.03 | 30.6±1.3 | 25.0±1.4 |
| Chelyabinsk (Lab. Measured)* | | | | | 27.9 | 22.8 |
| (8) Flora | 1.0±0.01 | 2.04±0.03 | 0.31±0.04 | 0.65±0.03 | 29.0±1.3 | 24.0±1.4 |
| (86039) 1999 NC43** | 0.957±0.01 | 1.989±0.03 | 0.8±0.04 | 0.53±0.03 | 23.2±1.3 | 19.6±1.4 |

*Laboratory measured fayalite and ferrosilite values for Chelyabinsk are from Meteoritical Bulletin Database. (http://www.lpi.usra.edu/meteor/metbull.php)
**BAR measured by de León et al. (2010) uses only half of the Band II area as originally described in Cloutis et al. (1986). The BAR from this work uses more modern definition, which includes the whole Band II area. Since olivine ratio is derived from BAR, the values are similarly affected.

**Figures**

Figure 1.

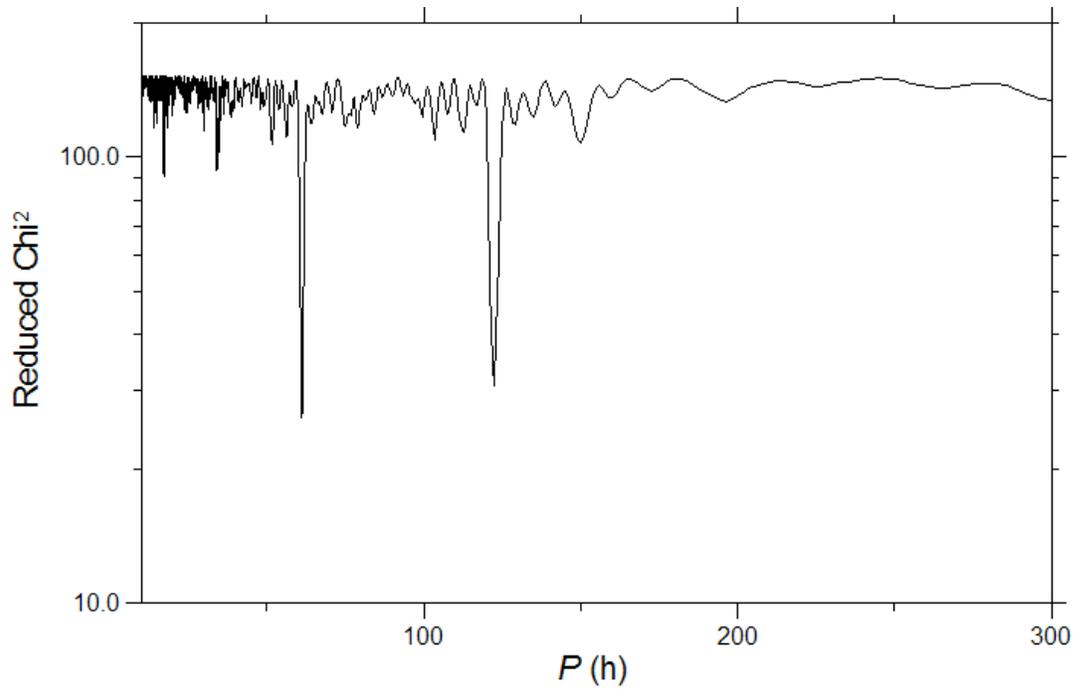

Figure 2.

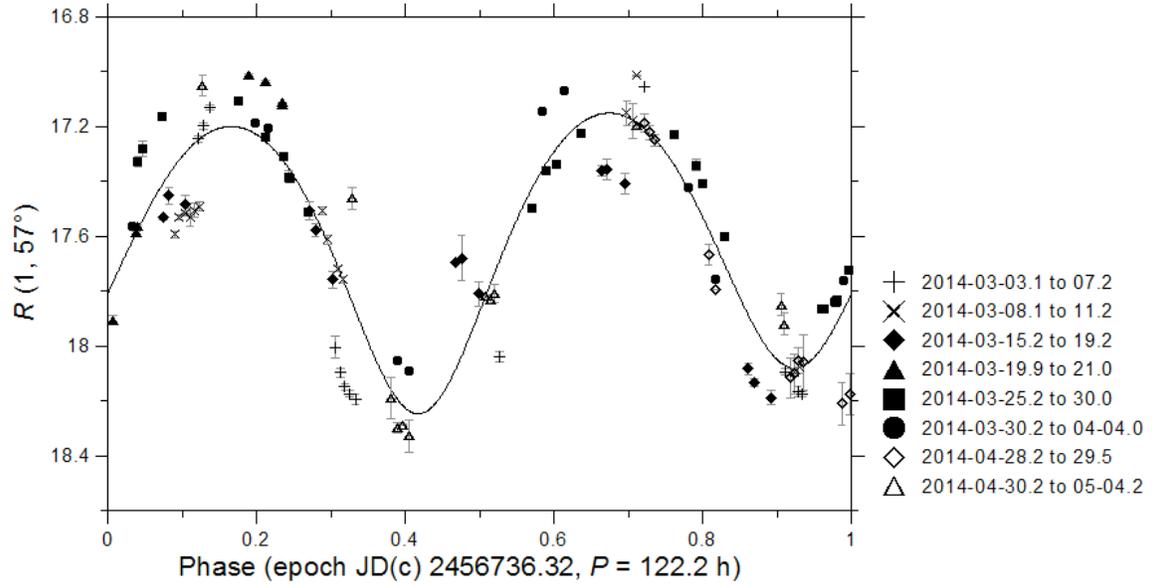

Figure 3.

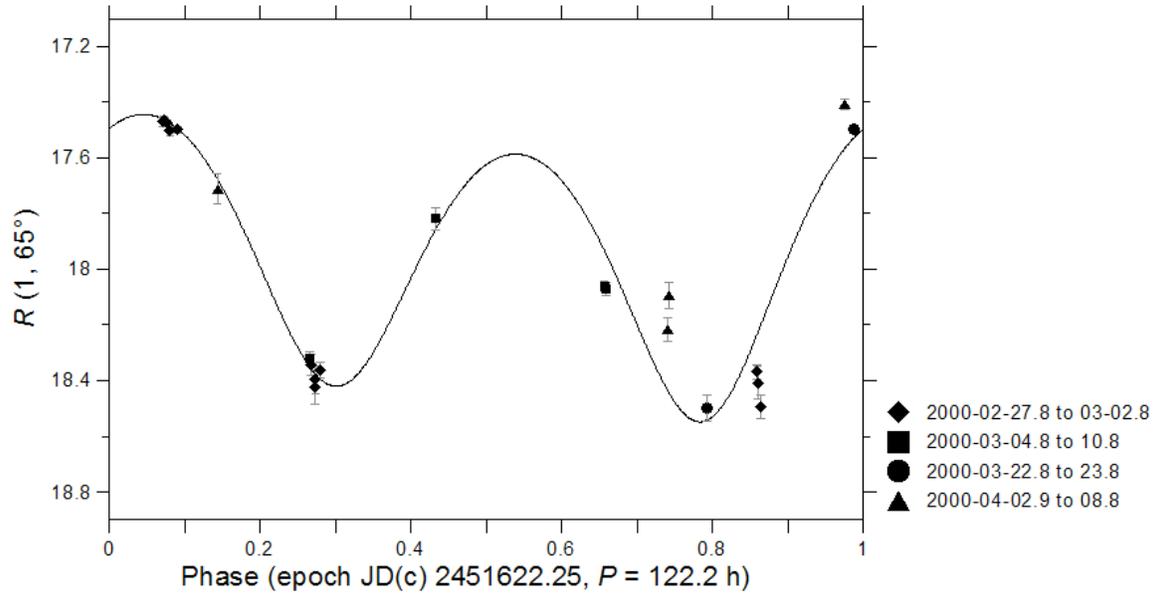

Figure 4.

Figure 5.

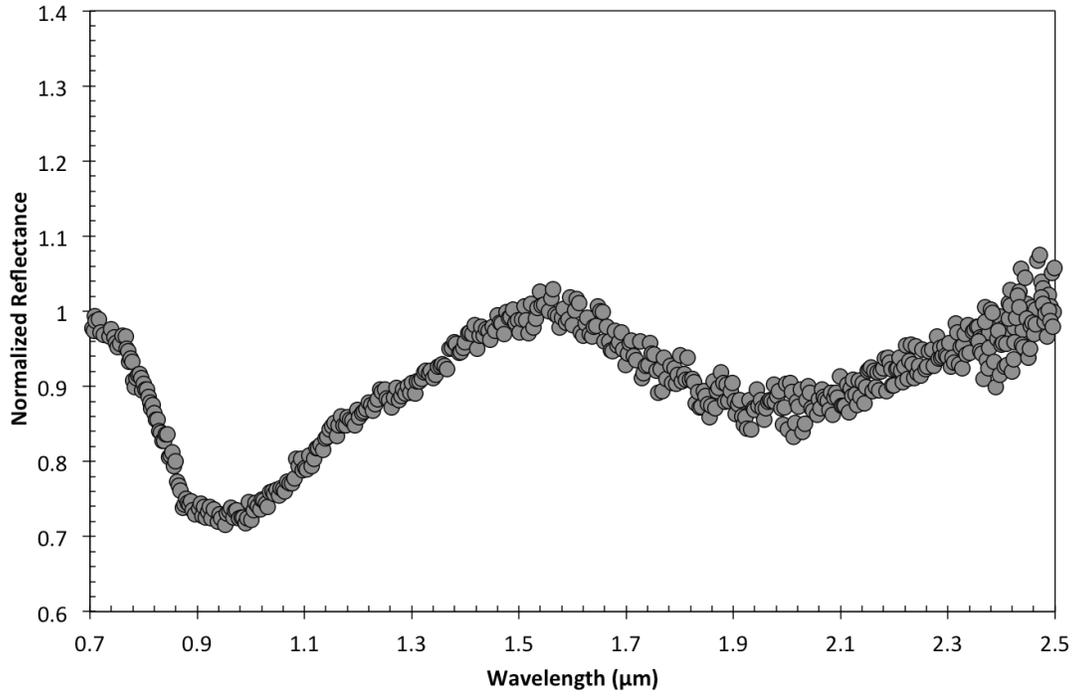

Figure 6.

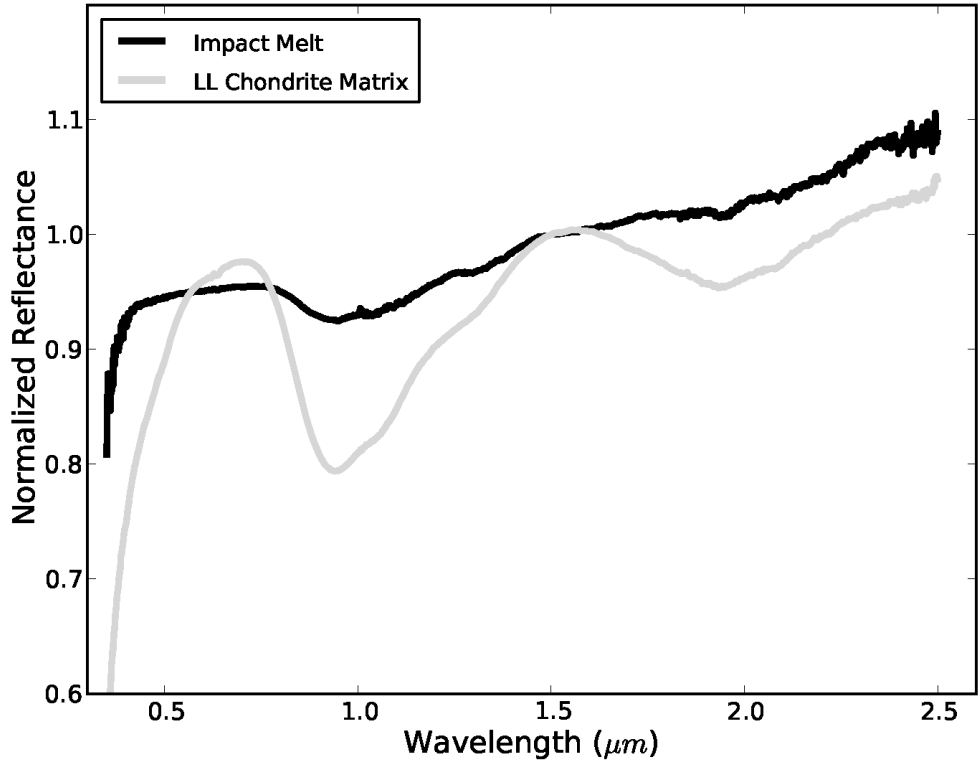

Figure 7.

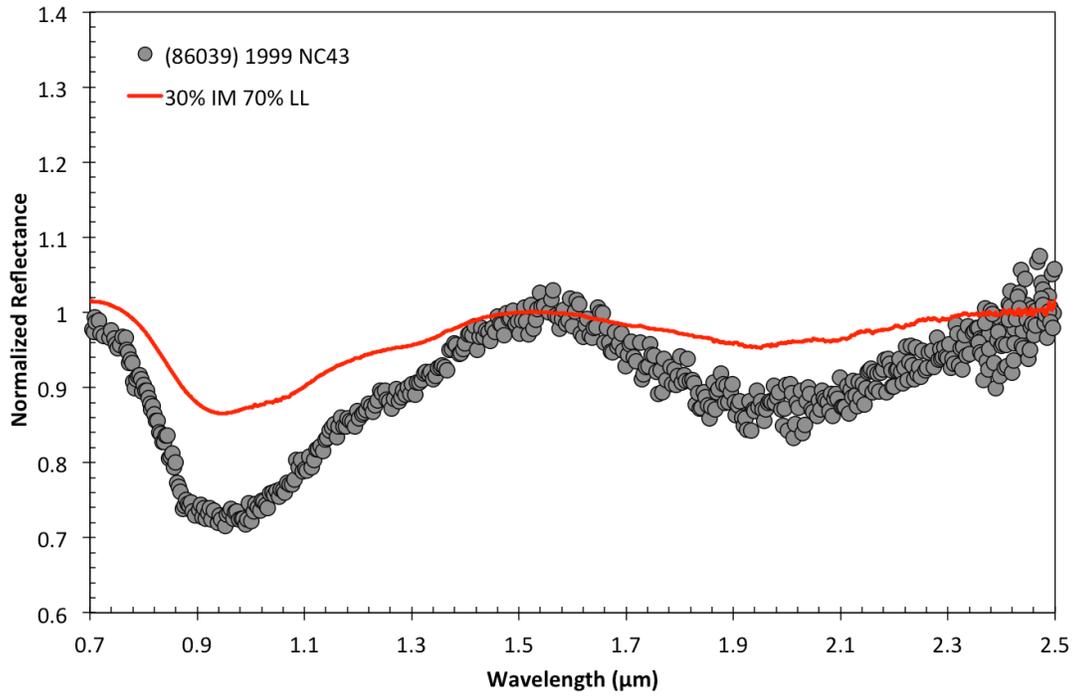

Figure 8.

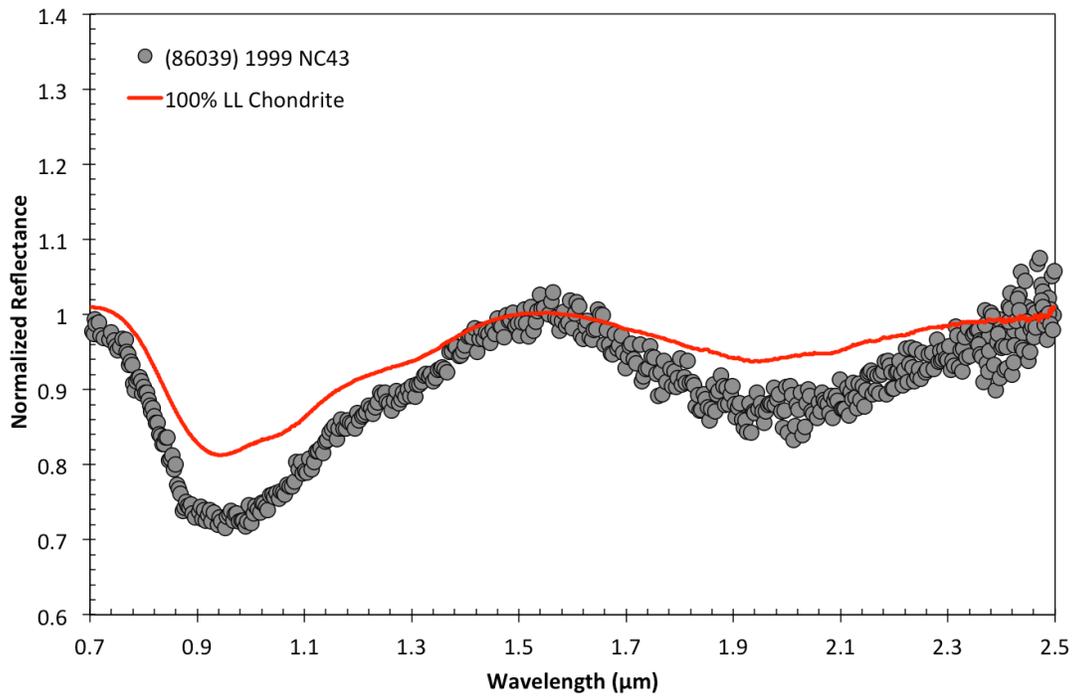

Figure 9.

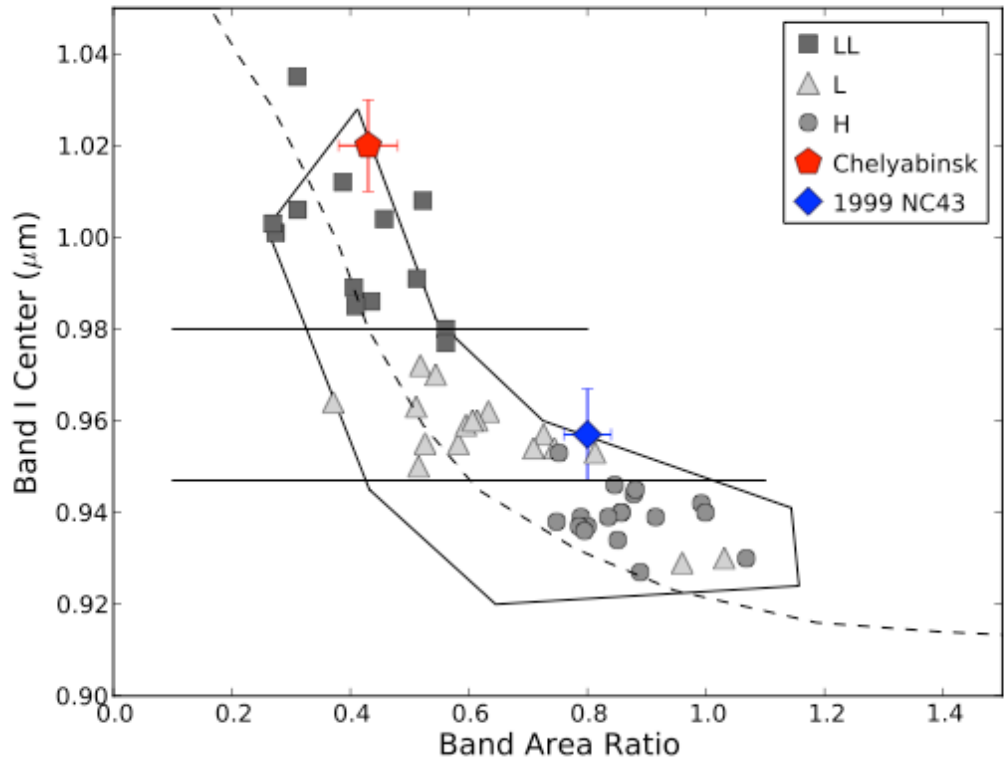

Figure 10.

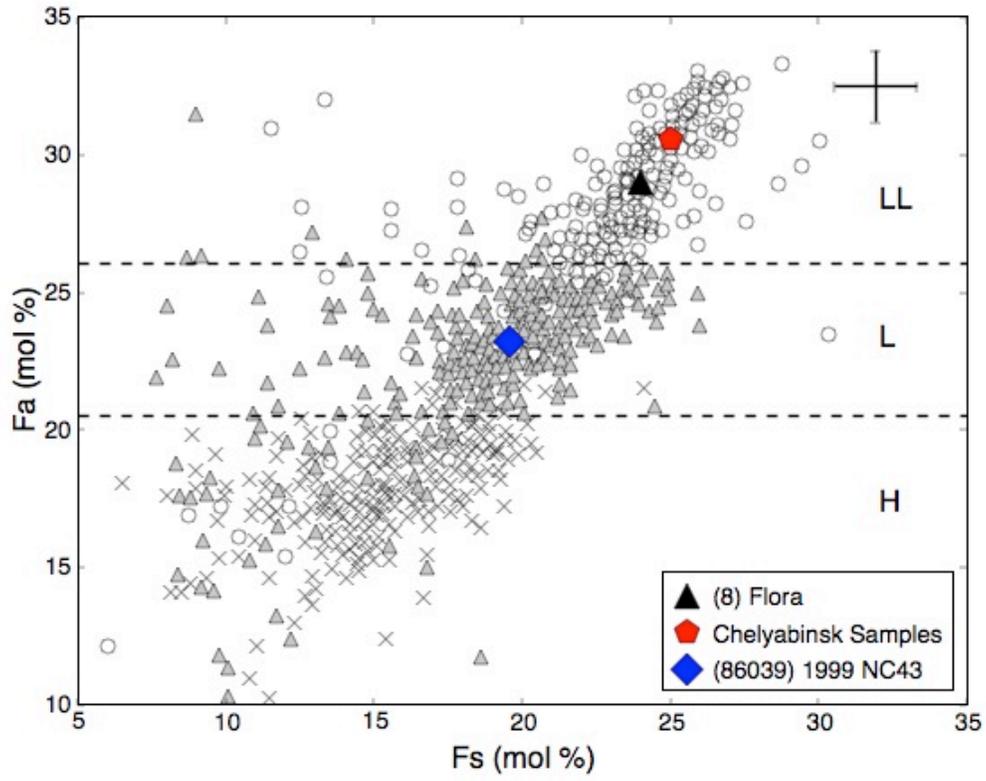

Figure 11.

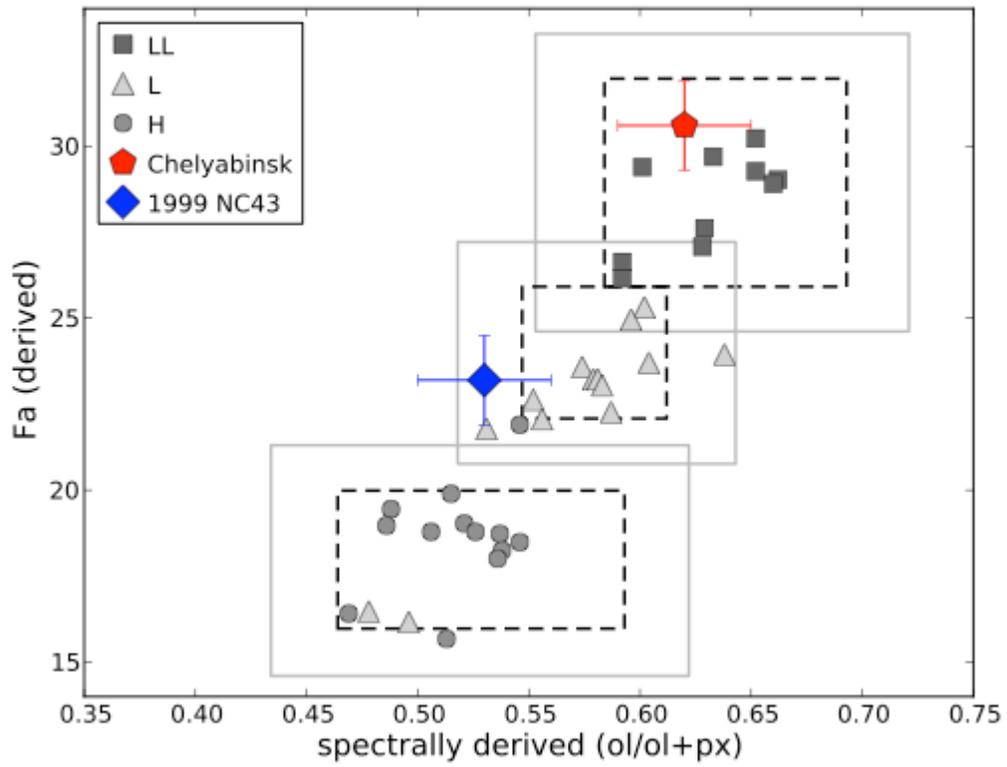

Figure 12.

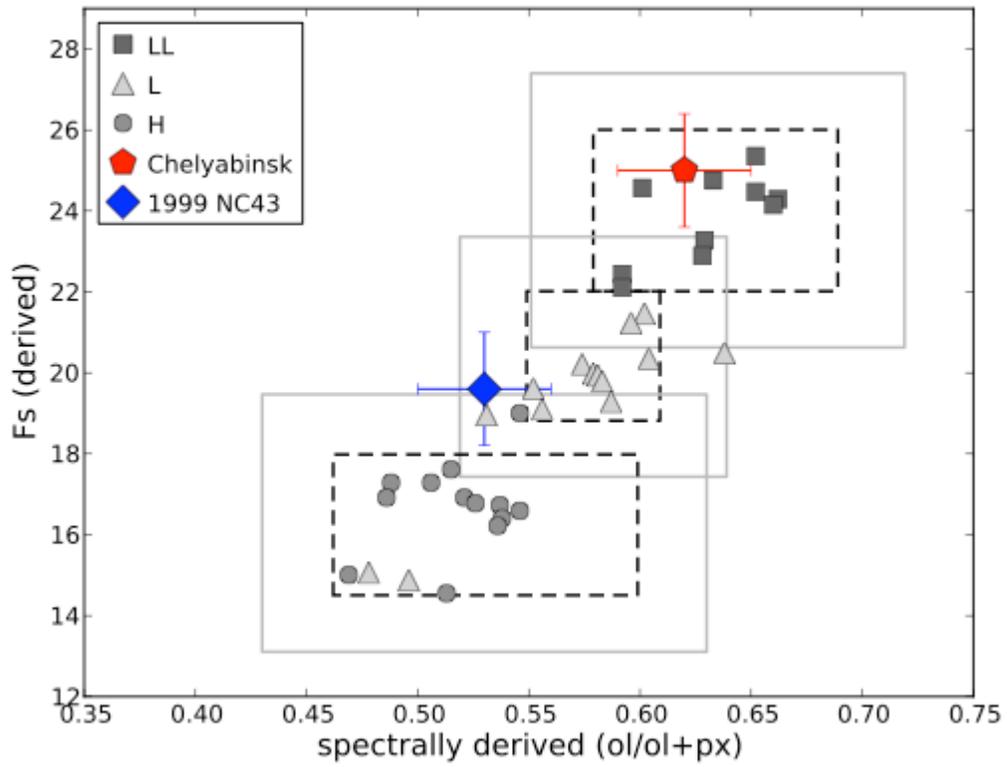

Figure 13.

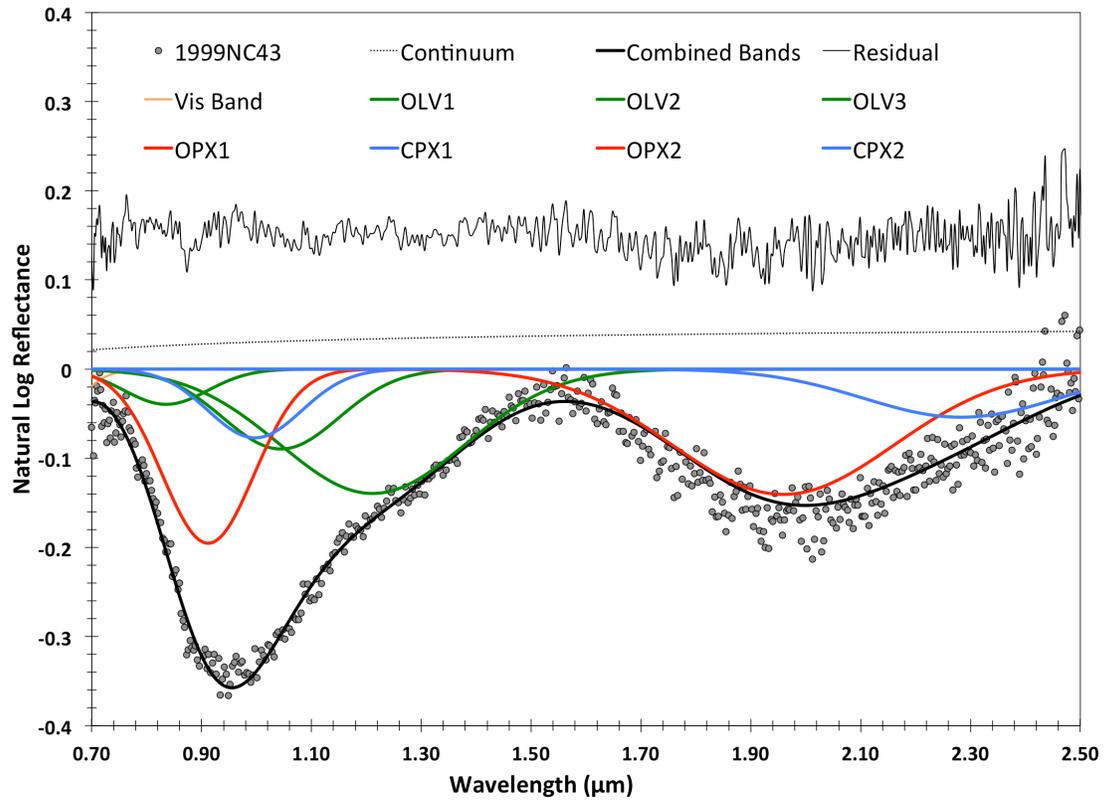